\documentclass[10pt,twocolumn,floatfix,footinbib,showpacs,preprintnumbers,amsmath,amssymb,pre,superscriptaddress,longbibliography]{revtex4-1}

\usepackage{epsfig,amscd,wasysym}
\usepackage{times}
\usepackage{amsmath}
\usepackage{amsfonts}
\usepackage{amssymb}
\usepackage{graphicx}
\usepackage{color}
\usepackage{accents}
\usepackage[margin=20mm]{geometry}
\usepackage{etex,array,mathtools,xcolor,physics,xr-hyper,hyperref,natbib}
\usepackage{subfigure}
\usepackage{dsfont}
\usepackage{todonotes}

\hypersetup{citecolor=violet}
\hypersetup{linkcolor=red}
\hypersetup{urlcolor=blue}
\hypersetup{colorlinks=true}

\newcommand{\be}{\begin{equation}}
\newcommand{\ee}{\end{equation}}
\newcommand{\bea}{\begin{eqnarray}}
\newcommand{\eea}{\end{eqnarray}}
\newcommand{\non}{\nonumber}
\newcommand{\Ket}[1]{     |   \,    #1    \rrangle}
\newcommand{\BraKet}[2]{\llangle #1 | #2\rrangle}
\newcommand{\Bra}[1]{  \llangle #1  \,  |}

\makeatletter 
\newsavebox{\@brx}
\newcommand{\llangle}[1][]{\savebox{\@brx}{\(\m@th{#1\langle}\)}%
  \mathopen{\copy\@brx\kern-0.5\wd\@brx\usebox{\@brx}}}
\newcommand{\rrangle}[1][]{\savebox{\@brx}{\(\m@th{#1\rangle}\)}%
  \mathclose{\copy\@brx\kern-0.5\wd\@brx\usebox{\@brx}}}
\makeatother

\newlength{\dhatheight} 

\newcommand{\qed}{\nobreak \ifvmode \relax \else
      \ifdim\lastskip<1.5em \hskip-\lastskip
      \hskip1.5em plus0em minus0.5em \fi \nobreak
      \vrule height0.75em width0.5em depth0.25em\fi}

\begin{document}
	
	\title{Effects of Quantum Pair Creation and Annihilation on a Classical Exclusion Process: the transverse XY model with TASEP}
	
	\author{K. Kavanagh\href{https://orcid.org/0000-0002-6046-8495}{\includegraphics[scale=0.066]{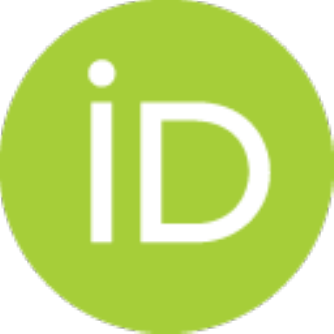}}}
	\email[Corresponding Author: ]{kevkav@stp.dias.ie}
	\affiliation{Dublin Institute for Advanced Studies, School of Theoretical Physics, 10 Burlington Road, Dublin 4, Ireland.}
	\affiliation{Department of Theoretical Physics, Maynooth University, Maynooth, Co. Kildare, Ireland.}
	
	\author{S. Dooley\href{https://orcid.org/0000-0002-2856-8840}{\includegraphics[scale=0.066]{ORCID.pdf}}}
	\affiliation{Dublin Institute for Advanced Studies, School of Theoretical Physics, 10 Burlington Road, Dublin 4, Ireland.}
			
	\author{J. K. Slingerland\href{https://orcid.org/0000-0002-9112-269X}{\includegraphics[scale=0.066]{ORCID.pdf}}}
	\affiliation{Department of Theoretical Physics, Maynooth University, Maynooth, Co. Kildare, Ireland.}
	\affiliation{Dublin Institute for Advanced Studies, School of Theoretical Physics, 10 Burlington Road, Dublin 4, Ireland.}
	
	\author{G. Kells \href{https://orcid.org/0000-0003-3008-8691}{\includegraphics[scale=0.066]{ORCID.pdf}}}
	\affiliation{Dublin City University, School of Physical Sciences, Glasnevin, Dublin 9, Ireland }
	\affiliation{Dublin Institute for Advanced Studies, School of Theoretical Physics, 10 Burlington Road, Dublin 4, Ireland.}

	\date{\today}
	
\begin{abstract}
	We investigate how particle pair creation and annihilation, within the quantum transverse XY model, affects the non-equilibrium steady state (NESS) and Liouvillian gap of the stochastic Totally Asymmetric Exclusion Process (TASEP). By utilising operator quantization we formulate a perturbative description of the NESS. Furthermore, we estimate the Liouvillian gap by exploiting a Majorana canonical basis as the basis of super-operators. In this manner we show that the Liouvillian gap can remain finite in the thermodynamic limit provided the XY model anisotropy parameter remains non-zero. Additionally, we show that the character of the gap with respect to the anisotropy parameter differs depending on the phase of the XY model. The change of character corresponds to the quantum phase transition of the XY model.
\end{abstract}

\preprint{DIAS-STP-21-14}	
\maketitle


\section*{Introduction}\label{sec:intro}
A valuable way of understanding a many-body system is to characterise its phase diagram and its associated transitions. This approach is useful across a broad class of domains, from the classical to the quantum realms, at zero-temperature, and both in- and out-of thermal equilibrium. Although typically such domains are clearly separated, there are situations where phase transitions in one such domain can influence the behaviour of another.

A useful framework  to address such issues is the Lindblad master equation \cite{Gorini1976, Lindblad1976}, through which one may combine both Hamiltonian and classical stochastic dynamics.  This methodology has been used, for example, to explore mixed classical-quantum transport \cite{Prosen2008, Prosen2008b, Eisler2011, Temme2012}. However, despite this success, it is difficult to find systems where an interesting interplay can be maintained between classical/stochastic and quantum phases. For example, for a spin chain with stochastic Lindblad processes only at the boundary spins, the typical steady state behaviour is dictated by the quantum properties of the bulk Hamiltonian (see e.g. \cite{Prosen2008,Prosen2008b}). On the other hand, if bulk stochastic processes are also allowed, these typically dominate \cite{Eisler2011, Temme2012} and leave little or no trace of the quantum phase transition to survive at late times.

In this paper we discuss a spin chain model where both classical stochastic and quantum phases are simultaneously relevant to a degree that allows for a genuine interplay between them in the long-time dynamics. The model is a combination of the transverse XY (TXY) Hamiltonian, or equivalently the Kitaev chain~\cite{Kitaev2001}, with a one-way classical stochastic hopping process, modelled by the Totally Asymmetric Simple Exclusion Process (TASEP). We refer to the combination of these two models as the TXY-TASEP.

The TASEP, considered in isolation, has a phase diagram for its non-equilibrium steady state (NESS) that is determined by the stochastic hop-on/hop-off rates at its boundaries. The TXY Hamiltonian undergoes a quantum phase transition in its ground state as the transverse magnetic field parameter is increased, assuming a non-zero XY anisotropy parameter $\delta$, at $\delta = 0$ the model is critical for any magnetic field value. When the two models are combined, we find that for zero anisotropy, $\delta=0$, the NESS retains many of the properties associated with the classical TASEP and, as such, its behaviour can be essentially controlled via the stochastic boundary (hop-on/hop-off) rates. On the other hand, in the regime associated with the anti-ferromagnetic (topological) phase of the XY model, the steady-state more closely resembles a perturbed infinite temperature state, but where the stochastic hop-on/hop-off rates do still dictate some key properties of the perturbation. 

The essential feature that allows for the balance between quantum and classical effects to be maintained is the non-zero XY anisotropy, which together with the bulk stochastic hopping, opens a constant Liouvillian gap that persists even for large system sizes. The precise scaling of the gap depends on the underlying quantum phase and is thus controlled by the bulk topology of the transverse XY model band-structure. This results in steady state properties that are very different in each of the quantum regimes.

From the perspective of the TASEP phase diagram \cite{Derrida1992}, we see that steady states of the low- and high-density phases are far more susceptible to the pair creation/annihilation associated with the XY anisotropy. This effect is much less pronounced in the maximal current phase, where the tendency of the XY anisotropy to drive the system towards half-filling is complementary to the maximal current micro-states.

Crucially, because of the constant gap, even in the thermodynamic limit one can move quickly between these limiting cases by simply tuning the transverse field. Systems with a finite gap in this limit are described as \emph{rapidly mixing} and it can be shown that the resultant steady states are robust to local perturbations and uncorrelated at a scale equivalent to the inverse gap size~\cite{Znidaric2015, Poulin2010, Nachtergaele2011, Kastoryano2013, Lucia2015, Cubitt2015}.  Our results, obtained by similar methods to prior studies of a dissipative XY model~\cite{Bardyn2012, Joshi2013}, suggest that the XY system parameters can be used to quickly engineer and tune specific features into the steady state and as such have the potential to be used as a means of rapid state preparation. 

The TXY-TASEP system does not allow for a direct analytical treatment, as available for related models \cite{Gwa1992, Kim1995, deGier2005, deGier2006, Prosen2008, Zunkovic2010, Crampe2010, Crampe2012, Lazarescu2014, Znidaric2015, Prolhac2016, Brattain2017, Zhang2019, Essler2020, Ishiguro2021, Robertson2021}. Our results are therefore arrived at by using a mix of numerical methods and approximate approaches. On a numerical level we apply matrix product state (MPS) methods~\cite{Nagy2002, Schollwock2011, Paeckel2019} to study steady states and the Liouvillian gap~\cite{Orus2008, Prosen2009, Joshi2013}. However, we also use operator quantization \cite{Prosen2008, Zunkovic2010}, and exploit the block structure that occurs naturally via the associated canonical Majorana representation \cite{Goldstein2012, Kells2015}, to make concrete perturbative statements. 
 
An overview of the paper is as follows:
In section~\ref{sect:model} we introduce key aspects of the transverse XY and TASEP models, providing in addition a detailed summary of our main results and the physical picture that emerges. In section~\ref{sect:NESS_results} we detail our main numerical results, focusing in particular on the relationship of the non-equilibrium steady state (NESS) with both the TASEP steady state and the maximally mixed state. In section~\ref{sect:gap_results} we discuss the Liouvillian super-operator of the model from the perspective of operator quantization and outline its block structure in what is called the canonical Majorana representation. This sets up our perturbative analysis of the NESS in the weak-stochastic limit \cite{Temme2012} and the subsequent focus on the two-quasiparticle super-operator block \cite{Prosen2008, Kells2015}. We provide a number of appendices for peripheral discussions. App.~\ref{app:discrete_TASEP} derives the continuous time TASEP master equation from the discrete time process. The remaining appendices (App.~\ref{app:BPT}, \ref{app:oddevengaps} \& \ref{app:oddspectrum}) expand on the technical aspects and interpretations of the block perturbation theory used in section~\ref{sect:gap_results}.

\section{Model and methods}\label{sect:model}

\subsection{Combining the TXY \& TASEP Models}
Our model of study is the combination of two paradigmatic models for transport in 1-dimensional systems: the  transverse-field XY model  (TXY) and the Totally Asymmetric Exclusion Process (TASEP). Separately both TXY and TASEP are well  understood;  the quantum XY spin model with a transverse magnetic field can be solved exactly by mapping to free fermion model with superconducting terms present due to the XY anisotropy. Likewise, the classical TASEP  is solvable in the sense that there is an ansatz solution for the NESS. Although this ansatz solution predates the tensor network concept, it takes the form of a matrix product state~\cite{Derrida1992}.  

\begin{figure}
\centering
\includegraphics[width=\columnwidth]{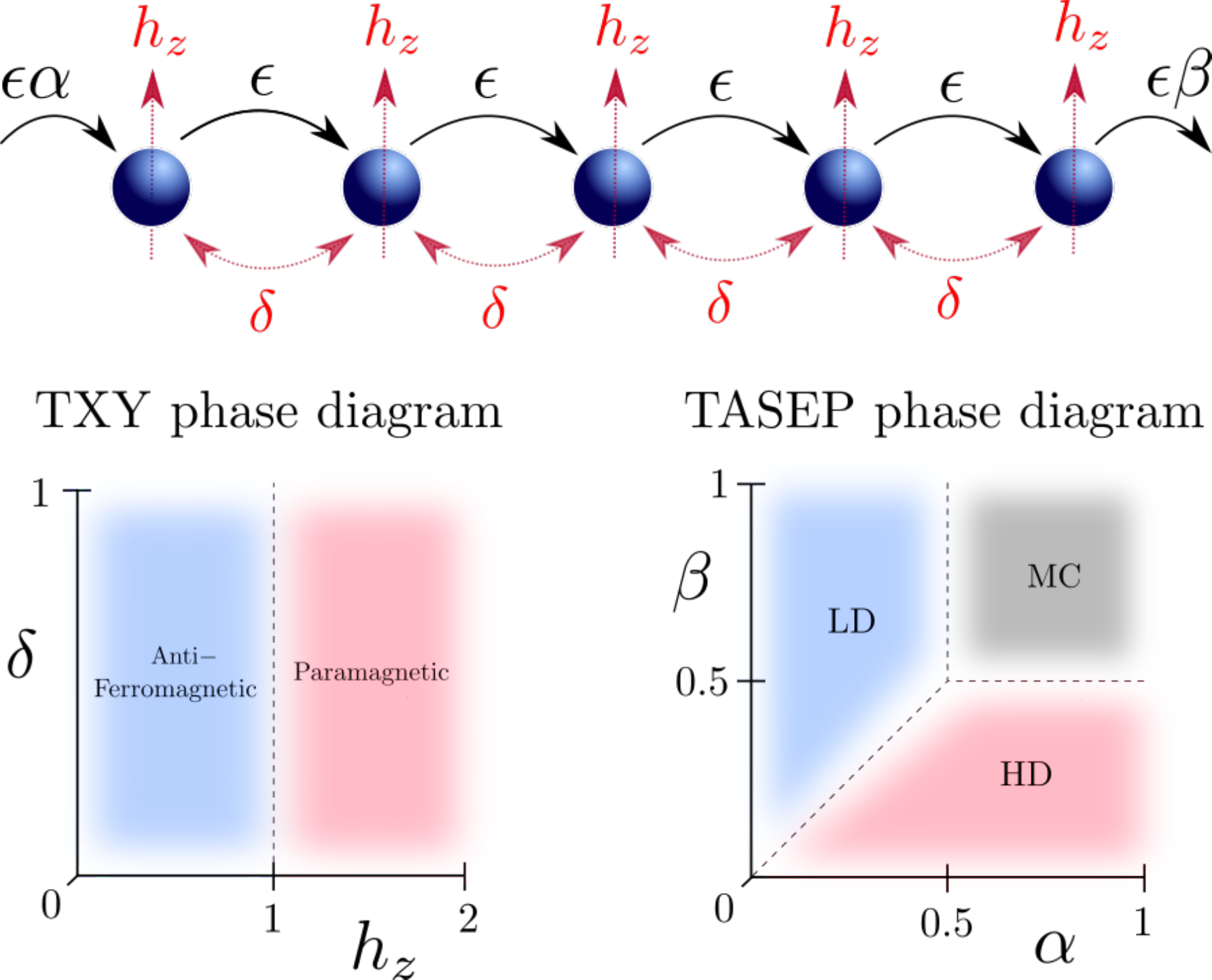}
\caption{\label{fig:model} Top: Our model is a chain of two-level quantum systems evolving by the combination of the Transverse XY Hamiltonian (TXY) and the totally asymmetric simple exclusion process (TASEP). The TXY model parameters are in red and TASEP parameters are in black. Bottom: the phase diagrams for the ground state of the TXY Hamiltonian (left), and for non-equilibrium steady state of the TASEP (right) where: LD = Low Density, HD = High Density, MC = Maximal Current.}
\end{figure}

We can incorporate both models into a single Lindblad master equation \cite{Gorini1976, Lindblad1976}
\begin{eqnarray} 
	 \frac{d\hat{\rho}}{dt} &=& - i \lambda \mathbb{H}(\hat{\rho}) + \epsilon \mathbb{L}(\hat{\rho}), \nonumber \\ 
	 &=& \mathcal{L}(\hat{\rho}). \label{eq:GKSL} 
\end{eqnarray}
The TXY model is represented by the following commutator \mbox{$\mathbb{H}(\rho) = [\hat{H}, \hat{\rho}]$}, with overall strength $\lambda$ and the Hamiltonian
\begin{equation} 
	\hat{H} = - h_{z} \sum_{j=1}^{N} \hat{\sigma}_{j}^{z} + \sum_{j=1}^{N-1} \left( \frac{1 + \delta}{2} \hat{\sigma}_{j}^{x}\hat{\sigma}_{j+1}^{x} + \frac{1 - \delta}{2} \hat{\sigma}_{j}^{y}\hat{\sigma}_{j+1}^{y} \right) . \label{eq:H} 
\end{equation}
Here $h_{z}$ is the transverse magnetic field and $0 \leq \delta \leq 1$ the anisotropy parameter. We note that if $\delta \neq 0$, the TXY-Hamiltonian has a quantum phase transition at $|h_{z}| = 1$ (see Fig. \ref{fig:model}). The anisotropic terms can be rewritten as $2\delta (\hat{\sigma}^{+}_{i}\hat{\sigma}^{+}_{i+1} + \hat{\sigma}^{-}_{i}\hat{\sigma}^{-}_{i+1})$, so they can be seen to introduce pair creation/annihilation when $\delta$ is non-zero. We make this statement in the view that, after a Jordan-Wigner transformation, $\hat{H}$ can be rewritten in terms of spinless fermions, which is known as the Kitaev chain \cite{Kitaev2001}. Then the spin model can be reinterpreted as particles hopping on a 1-dimensional lattice where spin-up corresponds to an occupied state and spin-down to an unoccupied state. 

In the second term of Eq.~\ref{eq:GKSL} we have the totally asymmetric simple exclusion process (TASEP), with overall strength $\epsilon$ and modelled by the Lindblad super-operator \cite{Temme2012}

\begin{equation} 
 \mathbb{L}(\hat{\rho}) = \alpha \mathcal{D}[\hat{\sigma}_{1}^{+}](\hat{\rho}) + \beta \mathcal{D}[\hat{\sigma}_{N}^{-}](\hat{\rho}) +\, \sum_{j=1}^{N-1} \mathcal{D}[\hat{\sigma}_{j}^{-} \hat{\sigma}_{j+1}^{+}](\hat{\rho}), \label{eq:cl_TASEP} 
\end{equation} 

\noindent where $\mathcal{D}[\hat{\ell}](\hat{\rho}) = \hat{\ell}\hat{\rho}\hat{\ell}^{\dagger} - \frac{1}{2}\hat{\ell}^{\dagger}\hat{\ell}\hat{\rho} - \frac{1}{2}\hat{\rho}\hat{\ell}^{\dagger}\hat{\ell}$. The TASEP is a classical stochastic process that involves hard-core particles hopping onto the first site of the chain with rate $\alpha$, hopping off the end of the chain with rate $\beta$, and hopping in one direction through the bulk with rate equal 1. The TASEP has three distinct phases with respect to $\alpha$ and $\beta$ (see Fig. \ref{fig:model}): the maximal current (MC) phase ($\alpha > 1/2$ and $\beta > 1/2$), the low density (LD) phase ($\alpha < 1/2$ and $\beta > \alpha$), and the high density (HD) phase ($\beta < 1/2$ and $\beta < \alpha$). This phase diagram can be deduced from an exact MPS solution for the TASEP steady state, with infinite dimensional matrices \cite{Derrida1992}. However, the exact solution can also be accurately approximated by a MPS with relatively small bond dimension \cite{Temme2012}. In this way we can generate an efficient matrix product state description of the TASEP steady state in a way that can be further extended to find the steady state $\hat{\rho}_\text{NESS}$ of the full Liouvillian $\mathcal{L}$, where an exact MPS is not known. Away from the purely classical model, we can obtain the full NESS by a density matrix renormalisation group (DMRG) implementation modified for open quantum systems~\cite{Orus2008, Prosen2009, Joshi2013}.

We note that Eq.~\ref{eq:cl_TASEP} is part of a continuous-time master equation, while TASEP is often considered as a discrete time stochastic process. In Appendix~\ref{app:discrete_TASEP} we outline the derivation of Eq.~\ref{eq:cl_TASEP} from the underlying discrete time stochastic process. By viewing the TASEP as a discrete time Markov process one can translate the model to a non-Hermitian spin chain for which Bethe anatz methods can be applied to determine analytic results, see e.g.~\cite{Gwa1992, Kim1995, deGier2005, deGier2006}. We note also that our approach is not the only one with the aim to introduce quantum effects into classical exclusion processes. A number of recent works have proposed quantum modified versions of the SSEP~\cite{Bernard2019, Bernard2021} and ASEP~\cite{Bernard2021sp} which employ a non-Hermitian Hamiltonian formulation of the exclusion process that introduces noise in the particle hopping amplitudes.

\begin{figure*}
	\centering
	\includegraphics[width=0.9\textwidth]{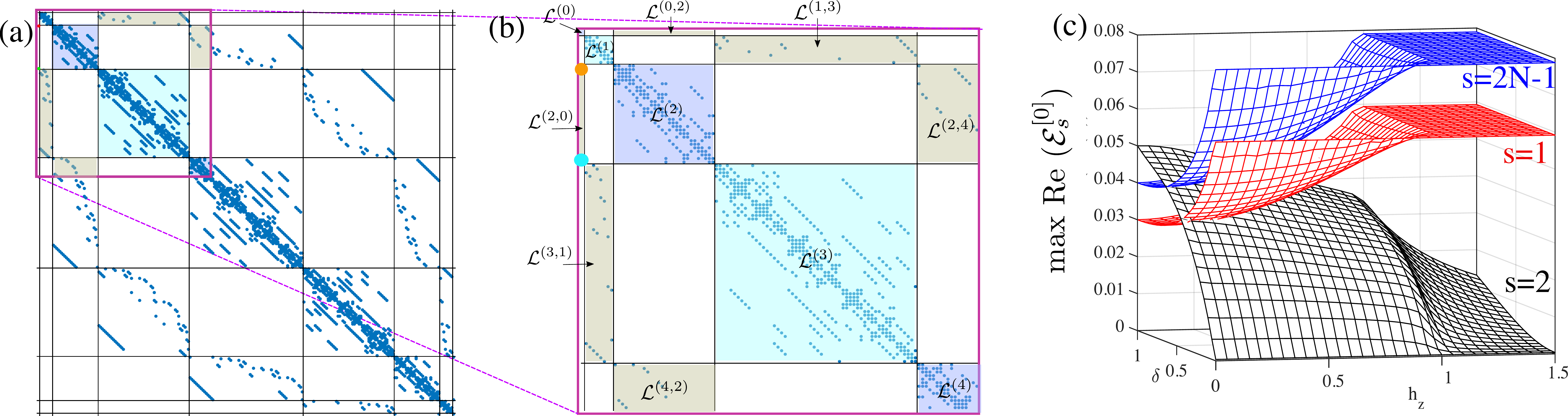}
	\caption{\label{fig:Lblocks}(Color Online) (a) The structure of $\mathcal{L}$ in the canonical basis for a system size of $N = 4$. (b) The $s = 0$ block that corresponds to the maximally mixed/thermal state is connected via terms dependent on the bulk and boundary driving to states $\Ket{\phi_L} = \Ket{\gamma_1 \gamma_2}$, $\Ket{\phi_R} = \Ket{\gamma_{N-1} \gamma_{N}}$. These elements are highlighted, on the left within the $\mathcal{L}^{(2,0)}$ sub-block, by the upper {\color{orange} \emph{orange}} dot and lower {\color{cyan} \emph{cyan}} dot, which have respective values $-\epsilon (\beta-1/2)$ and $ \epsilon (\alpha-1/2)$. (c)  One of our main observations is that the complex spectrum near $\mathcal{E} = 0$ is dominated by the states generated from the extremal blocks $\mathcal{L}^{(0)}$, $\mathcal{L}^{(1)}$, $\mathcal{L}^{(2)}$ and $\mathcal{L}^{(2N-1)}$ and that the eigenvalues of these states are well approximated by diagonalizing within each block separately. This can be seen via a non-Hermitian perturbative analysis where the effects of off-diagonal blocks appear only at second order, see Sec.~\ref{sect:perturbation_NESS}. In the figure, we give spectral gaps for $s = 1$ (red), $s = 2$ (black) and $s = 2N-1$ (blue) for a system of length $N = 100$, with $\alpha = 0.1$, $\beta = 0.3$, and $\epsilon = 0.1$.}
\end{figure*}

Our goal in this paper is to study the steady state and the Liouvillian gap of the corresponding TXY-TASEP model's Liouvillian super-operator $\mathcal{L}$, as we vary the model parameters, including the parameter $\epsilon/\lambda$ which controls the relative strength of the quantum TXY model and the classical TASEP in Eq.~\ref{eq:GKSL}. We set $\lambda = 1$ for the remainder of this paper, essentially allowing $\lambda$ to define the unit of frequency. We note that the steady state of $\mathcal{L}$ for the isotropic Hamiltonian, with $h_{z} = \delta = 0$ and TASEP, has been previously explored by other methods \cite{Temme2012}. Also, the case of zero bulk TASEP hopping has been explored in the more general scenario where particles can hop on or off either end of the chain \cite{Prosen2008b}. 

\subsection{Operator Quantization - A Hilbert Schmidt Formulation}\label{sect:OQ_basis}
\begin{figure*}
\centering
\includegraphics[width=1\textwidth]{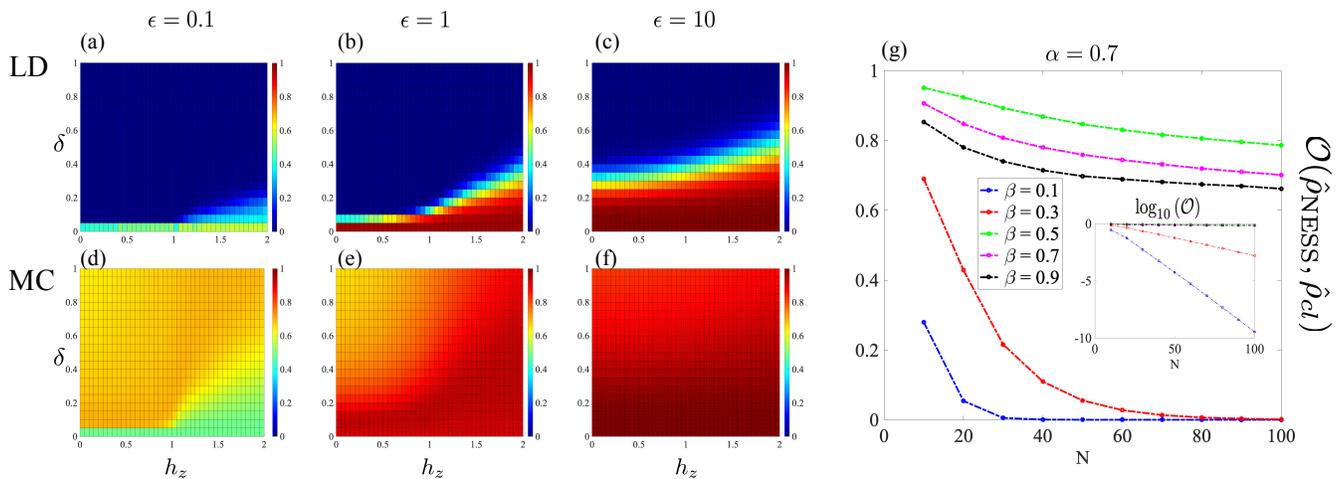}
\caption{\label{fig:New_Overlap_fig} (Color Online) These figures contain numerical data for the overlap, as defined in Eq.~\ref{eq:overlapDef}, for three cases of $\epsilon =\lbrace 0.1, 1, 10\rbrace$ and capturing features of the three TASEP phases. In (a)-(c), for the low density (LD) phase [$\alpha = 0.1$ \& $\beta = 0.3$] we observe a strong effect on the overlap with changing $\delta$, in the high density (HD) phase one can see similar features. In (d)-(f), for the maximal current (MC) phase [$\alpha = 0.7$ \& $\beta = 0.9$] we show the relatively weak effect of increasing $\delta$, note the restricted color range of values for this row of figures. In (g), the overlap is shown against system size, $N$, showing an exponential decay with system size within the high density (HD) phase (see inset showing $\log_{10}(\mathcal{O})$). In the MC phase the overlap decays at a slower rate with respect to system size. For (a)-(f), $N=50$. For (g), $\delta = 0.1, h_{z} = 0.5$ and $\epsilon = 0.1$.}
\end{figure*}

In the following, it will be useful to represent the superoperator $\mathcal{L}$ in Eq. \ref{eq:GKSL} as a matrix that acts on a vectorized representation of the quantum state $\hat{\rho}$. We do this by choosing a convenient basis of orthonormal operators $\{ \Gamma_i \}$ with respect to the Hilbert-Schmidt inner product, i.e., $\BraKet{\Gamma_i}{\Gamma_j} \equiv \Tr(\Gamma_i^\dagger \Gamma_j) = \delta_{i,j}$. We choose the so-called canonical Majorana basis \cite{Goldstein2012, Kells2015}: 

\begin{eqnarray}\label{eq:MFbasis}
	\Gamma^{(0)} : & \quad  & I/ \sqrt{2^{N}},\nonumber \\
	\Gamma^{(1)} : &  & \gamma_{1}/\sqrt{2^{N}},\gamma_{2}/\sqrt{2^{N}},\dots,\gamma_{2N}/ \sqrt{2^{N}},\\
	\Gamma^{(2)} :   &  & i\gamma_{1}\gamma_{2}/\sqrt{2^{N}},i\gamma_{1}\gamma_{3}/\sqrt{2^{N}},\dots,i\gamma_{2N}\gamma_{2N}/\sqrt{2^{N}},\nonumber \\
	& &\text{etc.}\non
\end{eqnarray}
These Majorana operators are defined from the spin operators as:
\begin{equation}
	\gamma_{2n-1} = \left(\prod^{2n-2}_{k=1}\sigma^{z}_{k}\right)\sigma^{x}_{2n-1}, \gamma_{2n}= \left(\prod^{2n-1}_{k=1}\sigma^{z}_{k}\right)\sigma^{y}_{2n},
\end{equation} 
for $n=1,2,\hdots,N$. As shown in Eq. \ref{eq:MFbasis}, an element $\Gamma^{(s)}_{a}$ of this basis is a product of Majorana operators, where the upper index $s$ is the number of $\gamma$'s in the product, and $a$ labels the basis elements within each $s$ subspace. The factors of $1/\sqrt{2^N}$ ensure the normalisation $\BraKet{\Gamma_a^{(s)}}{\Gamma_{b}^{s'}} = \delta_{s,s'}\delta_{a,b}$. In this basis the Liouvillian superoperator $\mathcal{L}$ has the matrix elements 
\begin{equation} 
	\mathcal{L}^{(s,s')}_{ab} = \llangle \Gamma^{(s)}_a |  \mathcal{L}(\Gamma^{(s')}_b) \rrangle,\label{eq:Labss}
\end{equation} 
where the upper indices $(s,s')$ label blocks in the matrix and the lower indices $a,b$ label the matrix elements within the $(s,s')$ block [see Fig. \ref{fig:Lblocks}(a,b) for an illustration of the matrix structure]. Likewise, the vectorized density operator in this operator basis has the vector elements $\rho_a^{(s)} = \Tr (\Gamma_a^{(s)}\rho)$.

The superoperator matrix $\mathcal{L}_{ab}^{(s,s')}$ can be non-Hermitian, resulting in a set of complex eigenvalues $\{ \mathcal{E}_0, \mathcal{E}_1, \mathcal{E}_2, \dots  \}$, which we assume are ordered according to their real parts $0 \geq \text{Re}(\mathcal{E}_0) \geq \text{Re}(\mathcal{E}_1) \geq \hdots$ etc.. The steady state corresponds to the eigenvalue with zero real part, $\text{Re}(\mathcal{E}_{0}) = 0$, and the Liouvillian gap is defined as
\begin{equation}
	\mathcal{E}_{gap} \equiv -\text{Re}(\mathcal{E}_{1}). \label{eq:L_gap}
\end{equation} 
The superoperator $\mathcal{L}$ has some other interesting features that are worth pointing out. First, we note that it preserves the parity of the label $s$ (i.e., the operator $\mathcal{L}(\Gamma^{(s)})$ is a linear combination of operator basis elements with the same parity as $s$). This is seen clearly in Fig. \ref{fig:Lblocks}(a,b), where $\mathcal{L}^{(s,s')} = 0$ if $s$ and $s'$ have different parity. Also, we highlight the $s=s'=0$ block [upper-left corner of Fig. \ref{fig:Lblocks}(a,b)], corresponding to the operator basis element $\Gamma^{(0)} = I/\sqrt{2^N}$. Using the master equation \eqref{eq:GKSL}, it is straighforward to show that this matrix element is always zero $\mathcal{L}^{(0,0)} = 0$. Similarly, it can be shown that this element is only connected to two others in the $\mathcal{L}^{(2,2)}$ block, via the off-diagonal block $\mathcal{L}^{(2,0)}$ [as illustrated in Fig. \ref{fig:Lblocks}(b)]. The two non zero elements are 
\begin{eqnarray} 
		  \llangle 2^{-\frac{N}{2}} \gamma_1 \gamma_2  | 2^{-\frac{N}{2}} I \rrangle &=& \epsilon(\alpha- 1/2),\nonumber \\  \BraKet{2^{-\frac{N}{2}} \gamma_{2N-1} \gamma_{2N}}{ 2^{-\frac{N}{2}} I} &=& -\epsilon(\beta- 1/2) . \nonumber 
\end{eqnarray}
	   If these two matrix elements are zero (i.e., if $\alpha=\beta=1/2$ or if $\epsilon = 0$) then the maximally mixed state $\rho \sim \Gamma^{(0)} \sim I$ is a valid steady state of the Liouvillian. If both matrix elements are non-zero but small then we expect the NESS to be close to the maximally mixed state. This intuition is based partly on the structure produced in our expression of the Liouvillian superoperator (see Fig.~\ref{fig:Lblocks} and Eq.~\ref{eq:Labss}) and on prior work for another system which allows for a NESS ansatz~\cite{Znidaric2011} with the maximally mixed state as the zeroth order state. We will exploit this feature later in Sections \ref{sect:perturbation_NESS} and \ref{sect:gap_results} to perturbatively estimate the steady state and the gap scaling in the small $\epsilon$ limit.
	  
Furthermore, generically speaking, for a Lindblad equation comprised of a Hamiltonian which is quadratic and Lindblad jump operators that are linear in fermion operators one finds that the Liouvillian super-operator admits a block diagonal matrix form. As a result, the super-operator can be solved block-by-block. There are cases however where exact treatments of the super-operator are possible despite the underlying Lindblad equation not being entirely quadratic. Asymmetric boundary driving~\cite{Prosen2008, Prosen2008b} and quartic stochastic processes~\cite{Eisler2011, Zunkovic2014} are two such examples. Although similar approaches cannot be directly applied to TXY-TASEP, we will show that using the canonical representation yields a useful block structure which allows for perturbative estimation of the Liouvillian gap in the weak classical regime.

\section{Non-Equilibrium Steady State}\label{sect:NESS_results}
The non-equilibrium steady state (NESS) is defined as the state $\hat{\rho}_\text{NESS}$ for which $\mathcal{L}(\hat{\rho}_\text{NESS}) = 0$.  The case for studying the NESS  is straightforward: it typically governs the system's late time behaviour. There are various examples of open quantum spin chains for which the NESS can be calculated exactly through analytical methods. One important class are those for which matrix product ansatz solutions exist for the NESS \cite{Znidaric2010c, Znidaric2011, Prosen2011b, Karevski2013, Prosen2015}. This includes, for example, the purely classical TASEP ($\lambda = 0$ in our model) for which a matrix product ansatz solution was found by Derrida \emph{et al.} \cite{Derrida1992}. Other formulations allow one to utilise the methodology from the Bethe Ansatz \cite{Gwa1992, Kim1995, deGier2005, deGier2006, Crampe2010, Crampe2012, Lazarescu2014, Prolhac2016, Brattain2017, Zhang2019, Essler2020, Ishiguro2021} or operator quantization \cite{Prosen2008, Prosen2008b, Eisler2011}. However, these exact analytical methods cannot be applied to the full TXY-TASEP to determine the NESS. Instead, in this section we employ the density matrix renormalisation group (DMRG) algorithm to numerically determine $\hat{\rho}_\text{NESS}$.

\subsection{Obtaining NESS from DMRG}
We begin by comparing $\hat{\rho}_\text{NESS}$ to the classical TASEP steady $\hat{\rho}_{cl}$, defined as the state for which $\mathbb{L}(\hat{\rho}_{cl}) = 0$ (where $\mathbb{L}$ is defined in Eq. \ref{eq:cl_TASEP}). For given TASEP boundary hopping rates $\left(\alpha, \beta\right)$ we know from the work of Derrida \emph{et al.}  \cite{Derrida1992} how to construct $\hat{\rho}_{cl}$ from its exact matrix product ansatz. However, introducing the Hamiltonian term in Eq.~\ref{eq:GKSL} typically modifies the steady state so that it is no longer equal to the classical TASEP steady state $\hat{\rho}_{cl}$. For a given $(\alpha, \beta)$ we quantify the difference between $\hat{\rho}_\text{NESS}$ and $\hat{\rho}_{cl}$ with the overlap

\begin{equation}\label{eq:overlapDef}
	\mathcal{O}(\hat{\rho}_\text{NESS}, \hat{\rho}_{cl}) = 
	\frac{\llangle \rho_\text{NESS} | \rho_{cl} \rrangle}{\sqrt{\llangle \rho_\text{NESS} | \rho_\text{NESS} \rrangle \llangle \rho_{cl} | \rho_{cl} \rrangle}}, 
\end{equation} 
	where $\llangle A | B \rrangle = \Tr (\hat{A}^\dagger \hat{B} )$ is the Hilbert-Schmidt inner product for operators $\hat{A}$ and $\hat{B}$. This overlap takes values in the interval $\mathcal{O} \in [0,1]$, with $\mathcal{O} = 1$ if $\hat{\rho}_\text{NESS} = \hat{\rho}_{cl}$ and $\mathcal{O} = 0$ if the states $\hat{\rho}_\text{NESS}$ and $\hat{\rho}_{cl}$ are orthogonal (i.e., $\llangle \rho_\text{NESS} | \rho_{cl} \rrangle = 0$).

Assuming $(\alpha, \beta)$ in the LD phase, in Fig. \ref{fig:New_Overlap_fig}~[(a)-(c)] we plot the overlap $\mathcal{O}$ as a function of the TXY-model parameters $(\delta,h_{z})$, for the three different TASEP strengths $\epsilon = \lbrace 0.1, 1, 10\rbrace$. For $\epsilon = 10$ the Liouvillian $\mathcal{L}$ is dominated by the TASEP component of the model. It is not surprising, therefore, that in Fig. \ref{fig:New_Overlap_fig}(c) we see large regions in parameter space where $\mathcal{O} \approx 1$. In particular, for small anisotropy $\delta$ we see that $\hat{\rho}_\text{NESS}$ and $\hat{\rho}_{cl}$ are very similar. This is consistent with previous work by Temme \emph{et al.} \cite{Temme2012}, which considered the transport properties for the TXY-TASEP in the special case of zero anisotropy $\delta = 0$, and found that the isotropic Hamiltonian has very little effect. However, even for $\epsilon = 10$ where TASEP dominates, we see in Fig. \ref{fig:New_Overlap_fig}(c) that increasing the TXY anisotropy parameter to relatively small values $\delta \gtrsim 0.5$ can lead to a significant decrease in the overlap $\mathcal{O}$. This suggests that, in the LD phase, the TXY anisotropy $\delta$ plays an important role in driving the NESS away from the TASEP steady state. Similar results are obtained for $(\alpha, \beta)$ chosen in the HD phase.

When $\epsilon  = 0.1$, on the other hand, the TASEP is relatively weak compared to the TXY Hamiltonian in Eq. \ref{eq:GKSL}, and so the steady state $\hat{\rho}_\text{NESS}$ may be very different from $\hat{\rho}_{cl}$. This is borne out in Fig.~\ref{fig:New_Overlap_fig}(a), where $\mathcal{O} \ll 1$ for most values of $(\delta,h_{z})$. However, even in this parameter regime we see a significant overlap $\mathcal{O}$ when $h_{z} > 1$ and $\delta$ is small, i.e., for parameters in the paramagnetic phase of the TXY-Hamiltonian. This indicates that the quantum phase transition affects the properties of the NESS.

\begin{figure}
\centering
\subfigure[~$\epsilon = 0.1$]{
\includegraphics[width=0.225\textwidth]{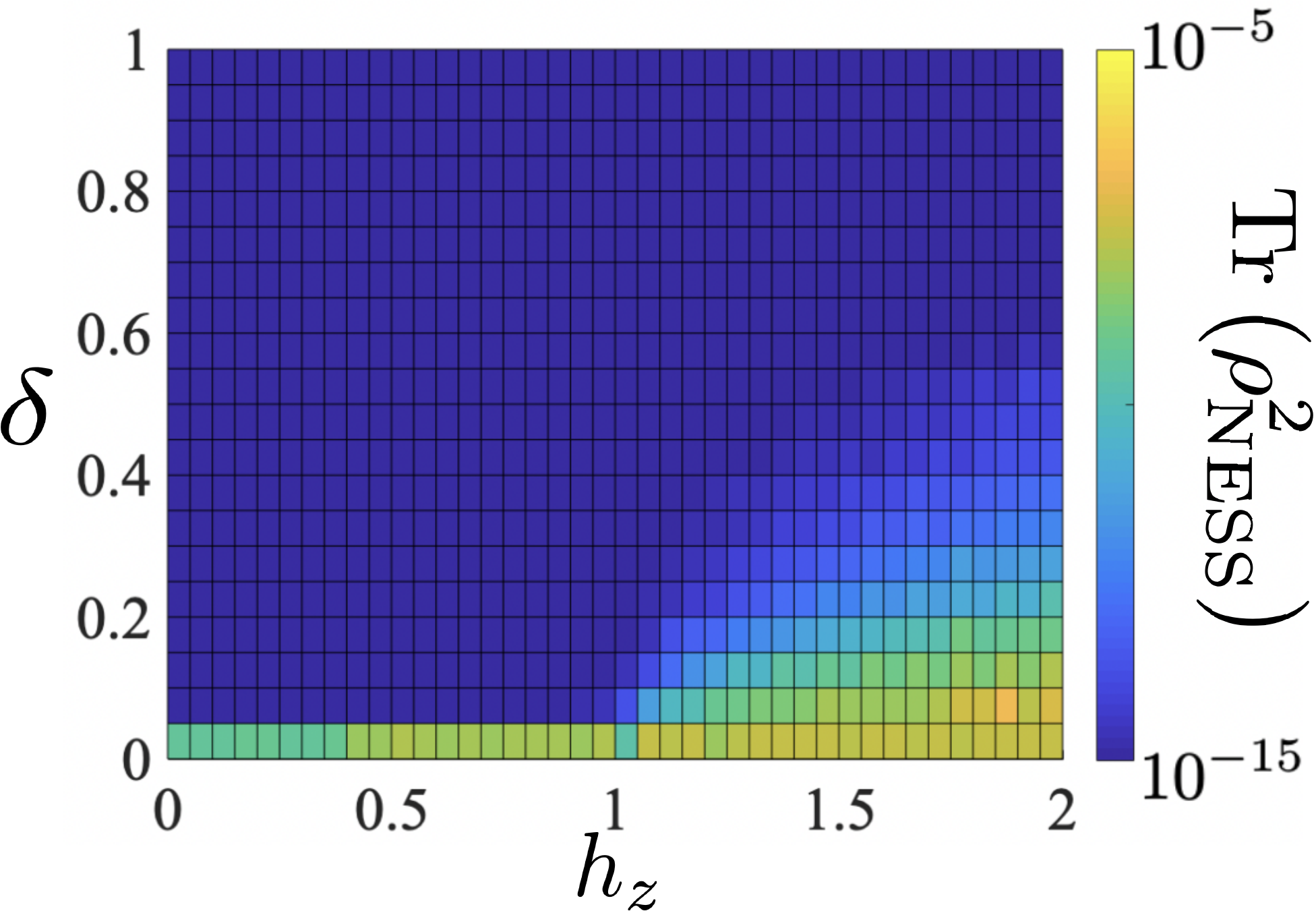}
}
\subfigure[~$\epsilon = 10$]{
\includegraphics[width=0.225\textwidth]{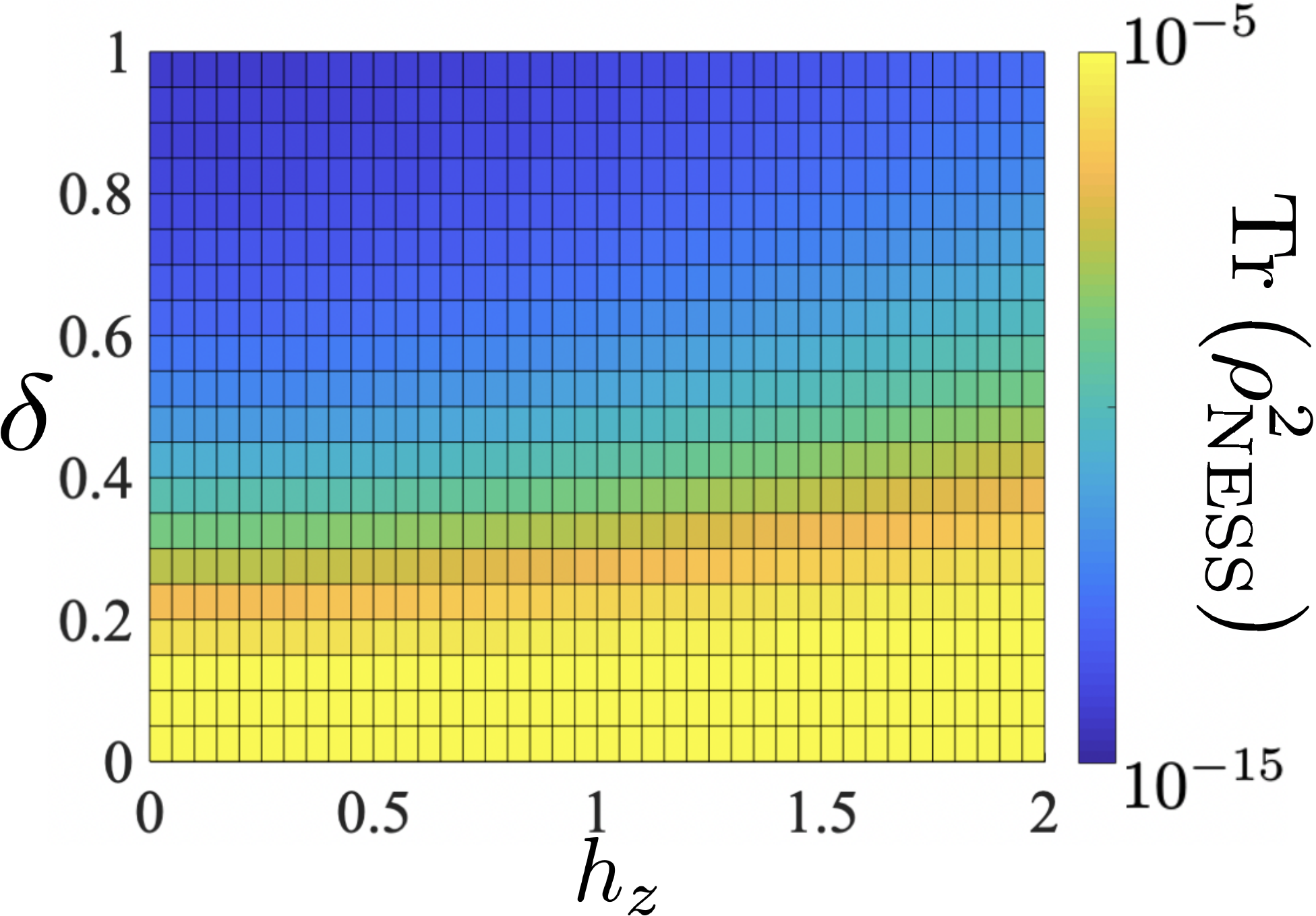}
}
\caption{\label{fig:mixednessLD} (Color Online) At classical rates $(\alpha,\beta)=(0.1,0.3)$, LD phase, we show the purity/mixedness of the NESS at two relative strengths $\epsilon$ representative of the weak/strong classical limits. In (a) $\epsilon = 0.1$, weak classical regime, we can see that increased $\delta$ quickly produces a more mixed state for all $h_{z}$ though more slowly for $h_{z} > 1$. In (b) $\epsilon = 10$, strong classical regime, the value of $h_{z}$ has less relevance yet the effect of increasing $\delta$ remains apparent. $N=50$ for both figures.}
\end{figure}

As mentioned above, our numerical results in Fig.~\ref{fig:New_Overlap_fig}~[(a)-(c)] are plotted for $(\alpha, \beta)$ in the LD phase, and similar results are obtained in the HD phase. However, the results are different for $(\alpha, \beta)$ in the MC phase. In Fig.~\ref{fig:New_Overlap_fig}[(d)-(f)] we can see that the overlap does not go to zero as in the LD phase for all $(\delta,h_{z})$. While an attempt has been made to highlight the different regions in $(\delta,h_{z})$, the overlap is largely similar across the parameter space. In Fig.~\ref{fig:New_Overlap_fig}(g) we plot the overlap $\mathcal{O}$ as a function of system size $N$, for various choices of $(\alpha, \beta)$. We see that the overlap decays much more slowly with system size for $(\alpha,\beta)$ in the MC phase.
								  
What can we say about $\hat{\rho}_\text{NESS}$ when it is driven away from $\hat{\rho}_{cl}$ by the TXY-Hamiltonian? We can gain some insight by studying the mixedness $\Tr(\hat{\rho}_\text{NESS}^2)$ of the steady state. In Fig.~\ref{fig:mixednessLD}[(a)-(b)] we plot the mixedness of the steady state $\hat{\rho}_\text{NESS}$ of the full Liouvillian in the LD regime. As the parameter $\epsilon$ decreases, corresponding to increasing relative strength of the TXY-Hamiltonian, we see that the NESS is driven away from $\hat{\rho}_{cl}$ to a much more mixed state. Moreover, with decreasing $\epsilon$ one can clearly resolve signatures of the phase transition of the XY-model at $h_{z} = 1$ and $\delta > 0$, see Fig.~\ref{fig:mixednessLD}(a).

\begin{figure}
\centering
\subfigure[~$N=50$, $\delta=0$]{
\includegraphics[width=.225\textwidth,height=0.2\textwidth]{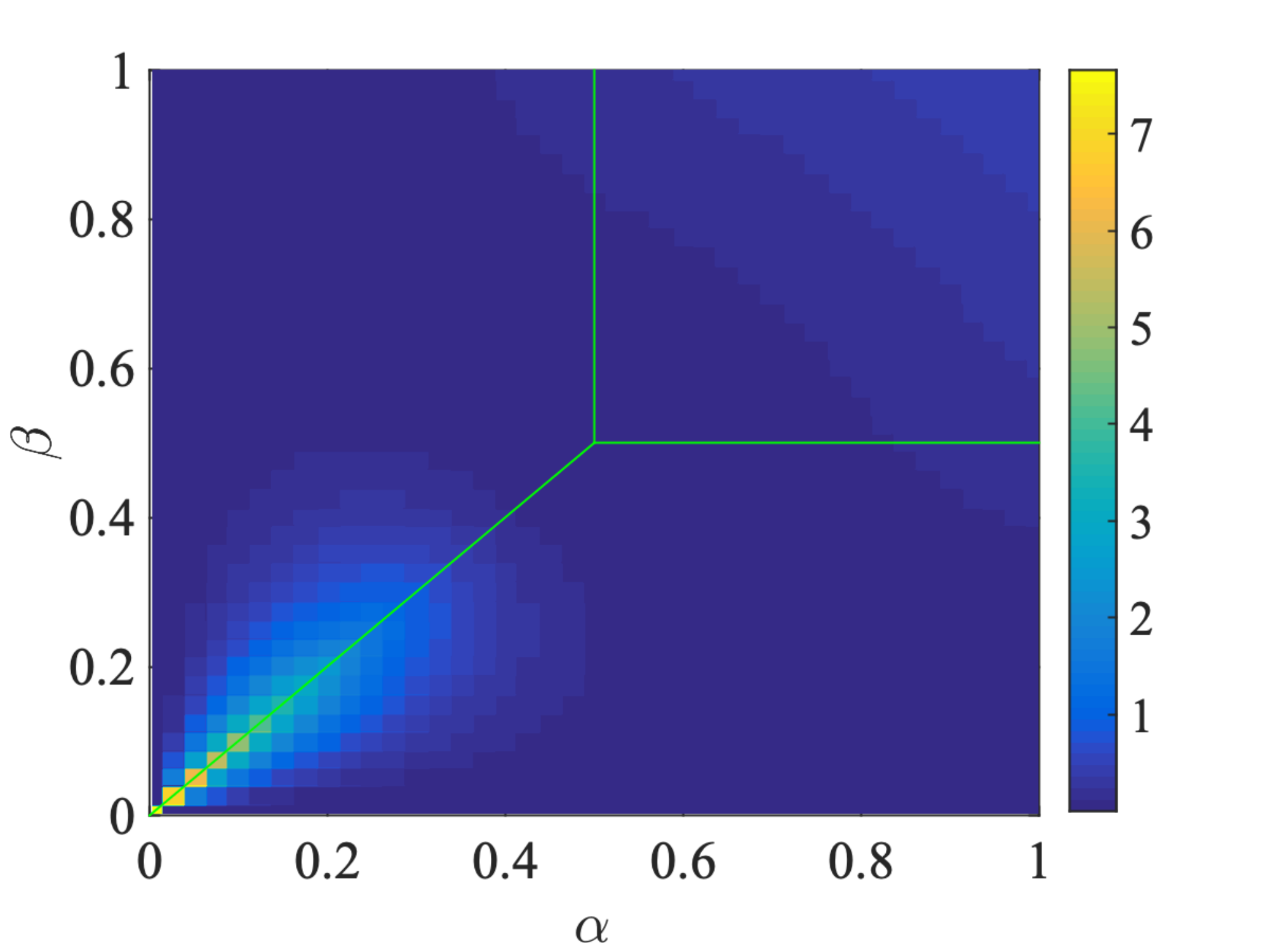}
\label{fig:KZ}
}
\subfigure[~$N=50$, $\delta=0.2$]{
\includegraphics[width=.225\textwidth,height=0.2\textwidth]{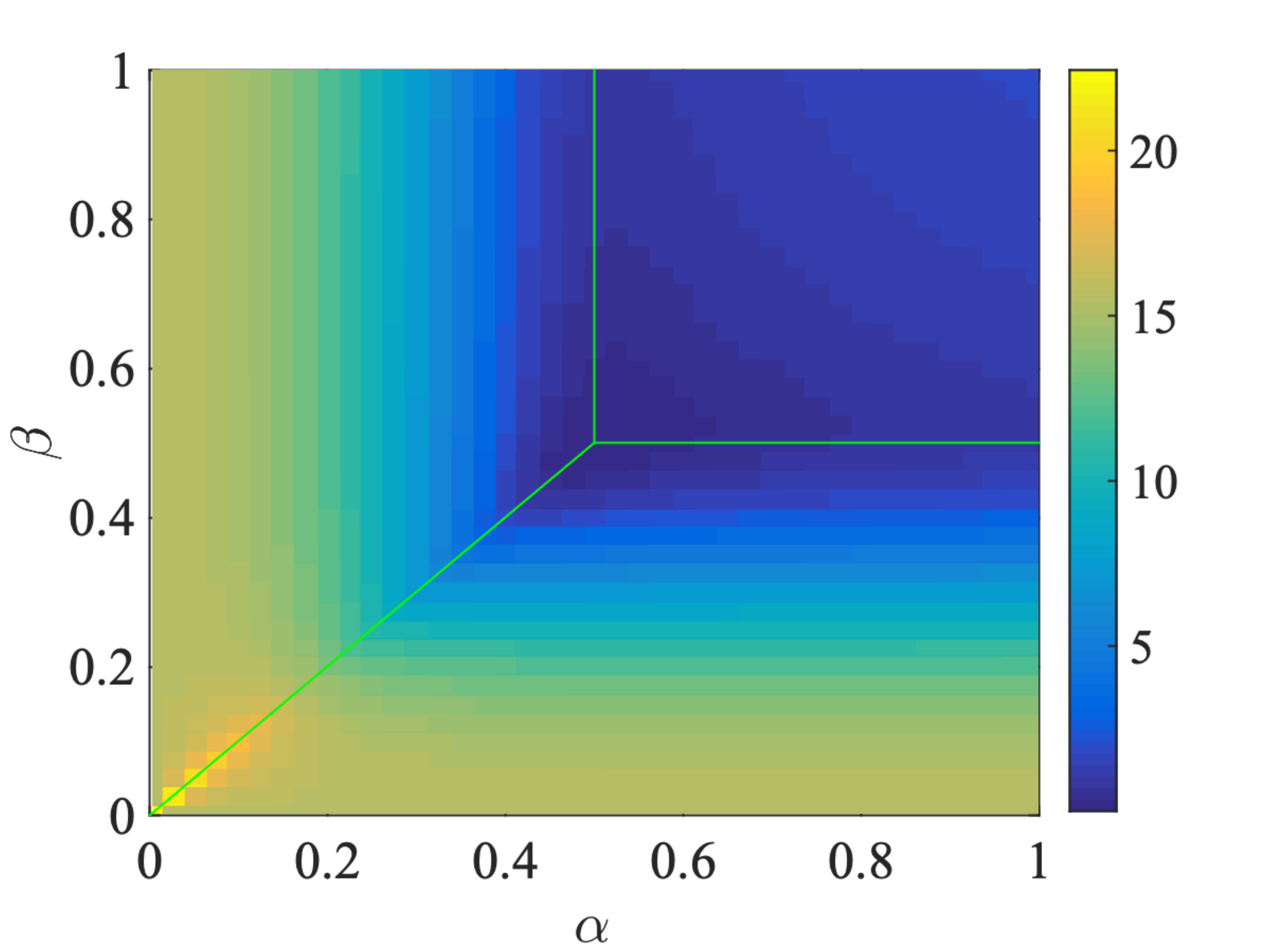}
\label{fig:KNZ}
}
\subfigure[~$\alpha = 0.7$, $\delta=0.1$]{
\includegraphics[width=0.45\textwidth]{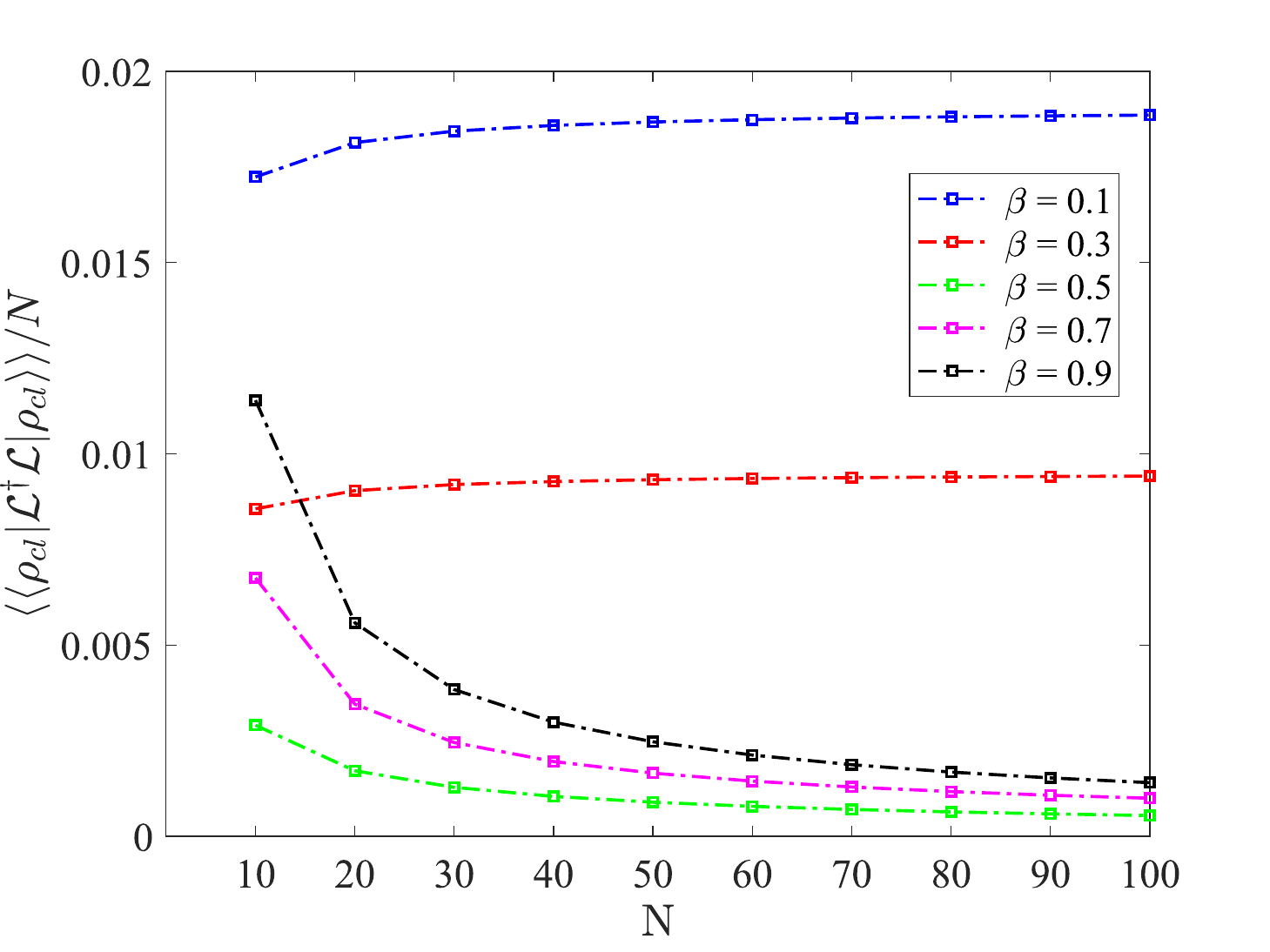}\label{fig:sus_cl_L_a7b19}}
\caption[]{\label{fig:commutator_expect} (Color online): The susceptibility of $\hat{\rho}_{cl}$ to dynamics by the Liouvillian $\, \Bra{\rho_{cl}} \mathcal{L}^\dagger \mathcal{L}\Ket{\rho_{cl}} = \lambda^{2}\Bra{\rho_{cl}} \mathcal{H}^2 \Ket{\rho_{cl}} $. [(a),(b)] The introduction of pairing $\delta$ allows the the classical steady state to couple strongly to the quantum commutator in both low and high density phases. (c) The strength of this coupling scales linearly with the system size in the low and high density phases (upper two lines). We emphasis this by plotting the susceptibility divided by $N$ so that the upper lines remain largely constant and the lower lines decrease. All data in this figure was plotted with $h_{z} = 0.5$ and $\epsilon = 0.1$.}
\end{figure}

We have shown then that increasing the TXY anisotropy can drive the NESS away from the classical TASEP steady state, for $(\alpha, \beta)$ in the LD/HD phase. To better understand this, we examine the overlap $\llangle \rho_{cl} | \mathcal{L}^\dagger \mathcal{L} | \rho_{cl} \rrangle = \big| \frac{d}{dt} |\rho_{cl} \rrangle \big|^2$, which quantifies the susceptibility of the TASEP steady state $\hat{\rho}_{cl}$ to dynamics of the full Liouvillian. Since $\mathbb{L}|\rho_{cl}\rrangle = 0$ we observe that $\llangle \rho_{cl} | \mathcal{L}^\dagger \mathcal{L} | \rho_{cl} \rrangle = \lambda^2 \llangle \rho_{cl} | \mathcal{H}^2 | \rho_{cl} \rrangle$, so that the susceptibility depends only on the Hamiltonian part of the Liouvillian. In Fig.~\ref{fig:commutator_expect}(a) we see that the isotropic Hamiltonian $\delta = 0$ has a relatively small effect on the classical steady state. However, for $\delta > 0$, Fig. \ref{fig:commutator_expect}(b) shows that $\hat{\rho}_{cl}$ responds very strongly to the TXY-Hamiltonian in the LD and HD phases, although not in the MC phase. This is reinforced by Fig. \ref{fig:sus_cl_L_a7b19}, which shows the susceptibility scales linearly with system size $N$ in the HD phase, but sub-linearly in the MC phase.

\begin{figure*}
\centering
\includegraphics[width=.95\textwidth]{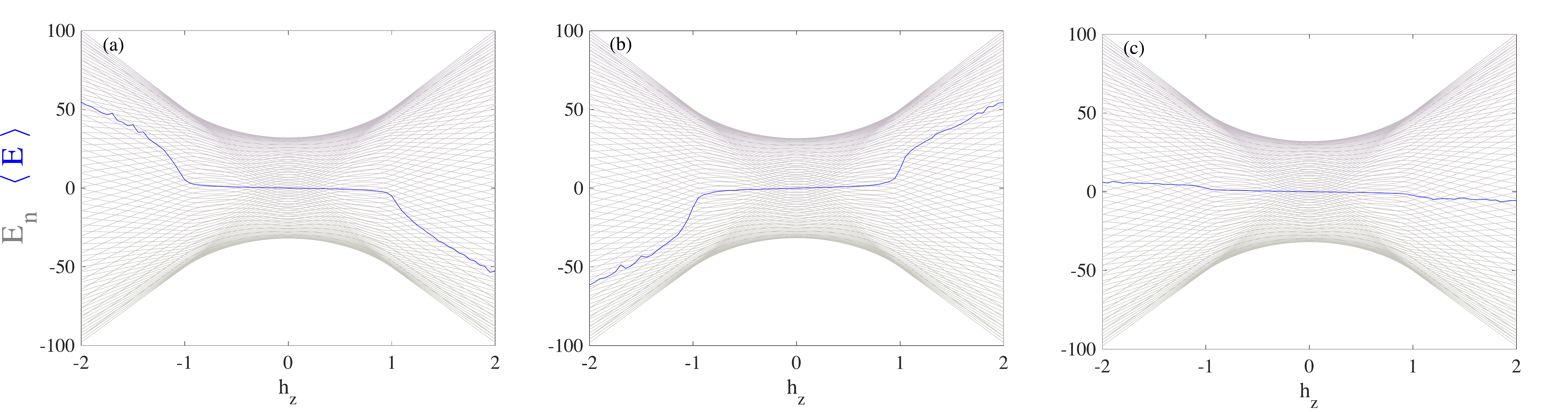}
\caption{\label{fig:Tr_rhoH}The energy expectation values of  $\langle E \rangle = \Tr\rho_{\text{NESS}}H$ together with the many-body eigen-spectrum, $E_n$ of $H$, where only band edges are shown. (a) $\alpha=0.1$, $\beta=0.3$, $\delta=0.1$, (b) $\alpha=0.3$, $\beta=0.1$, $\delta=0.05$, (c) $\alpha=0.7$, $\beta=0.9$, $\delta=0.1$. In all figures, $\epsilon  = 0.1$  and $N = 50$. In the paramagnetic regimes $|h_{z}| > 1$  the classical densities determined by the boundary rates result in steady states with a clear low/high energy imbalance [(a) and (b)]. This imbalance is suppressed in the ferromagnetic regime  $|h_{z}| < 1$ and also throughout the maximal current phase (c)}
\end{figure*}

One can intuit the reasons for the strong response of $\hat{\rho}_{cl}$ in this case by considering the steady state configuration \cite{Derrida1992, Rajewsky97, Evans1999, Nagy2002} in those classical phases. In the LD phase, as the name suggests, there are many empty sites. Rewriting the anisotropic terms of the Hamiltonian in Eq.~\ref{eq:H} as $\delta(\hat{\sigma}^{x}_{i}\hat{\sigma}^{x}_{i+1} - \hat{\sigma}^{y}_{i}\hat{\sigma}^{y}_{i+1}) = 2\delta (\hat{\sigma}^{+}_{i}\hat{\sigma}^{+}_{i+1} + \hat{\sigma}^{-}_{i}\hat{\sigma}^{-}_{i+1})$, the operator $\hat{\sigma}^{+}_{i}\hat{\sigma}^{+}_{i+1}$ associated with the anisotropy can successfully be applied to the state at many locations on the chain. Similarly, in the HD phase there are many occupied sites and the pair annihilation operator, $\hat{\sigma}^{-}_{i}\hat{\sigma}^{-}_{i+1}$, can be applied without annihilating the state. However, in the MC phase the steady state is largely comprised of half-filled configurations which will not couple as strongly to the anisotropic terms.

Another interesting property of the steady state $\rho_{\text{NESS}}$ is the expectation value $\langle E \rangle = \Tr \rho_{\text{NESS}} H$ which gives an indication of which Hamiltonian eigenstates take part in the steady state. In Fig.~\ref{fig:Tr_rhoH} we show how the expectation value changes relative to the full eigen-spectrum of the system Hamiltonian. In the LD and HD regimes the expectation value drifts towards the extrema of the the many body Hamiltonian spectrum, provided the Hamiltonian is tuned to the paramagnetic region. This occurs due to the energetic importance of either filled or empty sites (up or down spins) in this quantum phase. On the other hand, in the ferromagnetic/topological regimes, the energy of the steady state coincides with the centre of the many-body spectrum backing up the idea that here the system favours something close to the maximally mixed state. In the maximal current phase, this behaviour dominates for all values of the transverse field. 

\subsection{NESS as a Perturbation of the Maximally Mixed State}\label{sect:perturbation_NESS}
The perturbation theory utilised here for non-Hermitian systems is based on \cite{Sternheim1972,Li2014nat, Li2016}. For additional technical details see App.~\ref{app:BPT}. As a starting point, one defines a ``bare" unperturbed Liouvillian $\mathcal{L}_0$ with eigenvalues $\mathcal{E}_{n}$ and left and right eigenvectors $\Bra{\tilde{v}_{n}}$  and  $\Ket{v_{n}}$ such that $\Bra{\tilde{v}_m} \mathcal{L}_0 \Ket{v_n} = \delta_{nm} \mathcal{E}_n$. We write the perturbation as $\mathcal{L}_{1}$ and an expansion of the steady state as $\Ket{\rho} =\sum_{j} \Ket{\rho_{j}}$ the terms of which are produced iteratively according to 
\begin{equation}
	\Ket{\rho_{j}} =  \mathcal{L}_0^{-1} \mathcal{L}_1 \Ket{\rho_{j-1}},
	\label{eq:it_expand}
\end{equation}
where $\mathcal{L}_0^{-1}$ is the pseudo-inverse defined as
\begin{equation}
	\mathcal{L}_0^{-1} =\sum_{\mathcal{E}_n \neq 0}  \frac{\Ket{v_n} \Bra{\tilde{v}_n} }{ \mathcal{E}_n}.\label{eq:pseudo_inv}
\end{equation}
At this point one might expect that $\mathbb{H}$ is chosen as the unperturbed piece of the Liouvillian and subsequently that $\epsilon\mathbb{L}$ becomes the perturbation. However, one can immediately see an obstacle arising from this choice. Given Eq.~\ref{eq:pseudo_inv}, since $\mathbb{H}$ corresponds to the commutator of the Hamiltonian its spectrum is massively degenerate and all eigenvectors of the Hamiltonian yield zero eigenvalue in the commutator. As a result we would be left with a highly degenerate situation that is difficult to deal with.

We propose a way to circumvent this obstacle by exploiting the structure of $\epsilon\mathbb{L}$. We know that once any part of $\mathbb{L}$ is switched on that the system will immediately have a preferred steady state. As such we propose that to proceed we first treat diagonal ($\backslash$) components of the TASEP term $\mathbb{L}$ differently from the off diagonal ($\backslash \backslash$) ones. Namely, we split the total $\mathbb{L}$ as the sum
\begin{equation}
	\epsilon \mathbb{L}  \rightarrow  \epsilon_{\backslash}  \mathbb{L}_{\backslash}  +  \epsilon_{\backslash \backslash}  \mathbb{L}_{\backslash \backslash}.
\end{equation}
We note here that this expression of the splitting of $\epsilon\mathbb{L}$ is an equality, however we introduce new $\epsilon$ variables for the separate components for this calculation. In the end they will equalized to the original $\epsilon$ variable. Our unperturbed system will then consist of the collective diagonal blocks
\begin{equation}
	\mathcal{L}_0 = \sum_{ s  \in even}  \mathcal{L}^{(s)} =  \sum^{2N}_{s  \in \text{ even}}  \epsilon_{\backslash}   \mathbb{L}^{(s)}_{\backslash} -i \lambda \mathbb{H}^{(s)},
\end{equation}
and the perturbation as the remaining off diagonal components
\begin{align}
	\mathcal{L}_1 = \mathcal{L}-\mathcal{L}_0 &=&  \sum_{s  \in \text{ even}}   \mathcal{L}^{(s,s+2)} + \mathcal{L}^{(s+2,s)},\nonumber \\
	&=&  \epsilon_{\backslash \backslash } \sum_{s  \in \text{ even}}   \mathbb{L}^{(s,s+2)} + \mathbb{L}^{(s+2,s)}.
\end{align}
The block diagonal form of $\mathcal{L}_0$ means that we can write down its eigen-spectrum block by block. In practice we observe numerically that the real component of $\mathcal{E}_{n}^{(s)}$ for small $\epsilon_{\backslash}$ grows linearly such that in what follows it will be useful to write this dependence explicitly and expand the complex eigenvalue as $\mathcal{E}_n^{(s)} = \epsilon_{\backslash} r_{n}^{(s)} +i E_{n}^{s} $. 

Another property of our unperturbed operator is that the pseudo-inverses of the blocks only act locally within each block. This will allow us to simplify some expressions in the following and implies for example that
\begin{equation}
	\mathcal{L}_0^{-1} = \sum_{ s  \in even}  [\mathcal{L}^{(s)}  ]^{-1}.
\end{equation}

Then, with the maximally mixed state as our starting state $\Ket{\rho_0} =\Ket{I}$ we can proceed according to the iterative procedure \eqref{eq:it_expand}:
\begin{align}
	\Ket{\rho_1} &=& -  [\mathcal{L}^{(2)}]^{-1}   \mathcal{L}^{(2,0)}  \Ket{I}, \\
	\Ket{\rho_2} &=& - [\mathcal{L}^{(4)}]^{-1}   \mathcal{L}^{(4,2)}  \Ket{\rho_1},  \nonumber \\
	\Ket{\rho_3} &=& -  ([\mathcal{L}^{(2)}]^{-1}    \mathcal{L}^{(2,4)}  +   [\mathcal{L}^{(6)}]^{-1}   \mathcal{L}^{(6,4)}  ) \Ket{\rho_2}, \nonumber \\
	&\vdots & \non
\end{align}
where only the non-zero $\mathcal{L}^{(s,s^{\prime})}$ blocks/elements have been kept. Plugging in the dependence on the overall weights we have for the first order expression:
\begin{align}
	\Ket{\rho_1} 
	&=& -   \epsilon_{\backslash \backslash}  [\mathcal{L}^{(2)}]^{-1}   \mathbb{L}^{(2,0)}  \Ket{I}, \nonumber \\
	&=& -  \epsilon_{\backslash \backslash}   \sum_{n }  \frac{ \Ket{v_n^{(2)}}}{\mathcal{E}_n^{(2)}} \Bra{\tilde{v}^{(2)}_n}  \mathbb{L}^{(2,0)}  \Ket{I}, \nonumber \\
	&=& -  \epsilon_{\backslash \backslash}   \sum_{n }  \frac{  \bar{\alpha}  \BraKet{\tilde{v}^{(2)}_n}{\phi_L} - \bar{\beta}  \BraKet{\tilde{v}^{(2)}_n}{\phi_R} }{\epsilon_{\backslash} r^{(2)}_{n} +  i E^{(2)}_{n}   }    \Ket{v_{n}},
\end{align}
with $\Ket{\phi_L} = \Ket{\gamma_1 \gamma_2}$, $\Ket{\phi_R} = \Ket{\gamma_{2N-1} \gamma_{2N}}$, $\bar{\alpha} = \alpha -1/2$, $\bar{\beta} = \beta -1/2$  and where on the last line we have also expanded the $s=2$ block eigenvalues into their real and imaginary components.
 
Leaving the inner products in the numerator to one side for a moment we can consider which terms are relevant in this first iterative correction by looking at cases for the coefficients in the sum: 
\begin{equation}\label{eq:firstordercorr}
\frac{-\epsilon_{\backslash \backslash}}{\epsilon_{\backslash} r^{(2)}_{n} +  i E^{(2)}_{n}} \sim
\begin{cases}
	\frac{-1}{\,r^{(2)}_{n}},\, E^{(2)}_{n} \ll \epsilon_{\backslash} r^{(2)}_{n},\\
	\frac{i \epsilon_{\backslash}}{E^{(2)}_{n}}, \text{otherwise.}\\
\end{cases}
\end{equation}
Evidently as we reinstate $\epsilon_{\backslash \backslash} = \epsilon_{\backslash} \rightarrow \epsilon$ and approach $\epsilon \rightarrow 0$ the second case is irrelevant and only those coefficients with small to negligible imaginary parts contribute to the correction.

What about the terms  $\BraKet{\tilde{v}^{(2)}_n}{\phi_{L/R}}$? An unusual feature of the block-decomposition is that we could in principle have additional $\epsilon_{\backslash}$ dependences occurring through the $\Ket{ \tilde{v}^{(2)}_n}$. However, in practice we see via direct evaluation that, to leading order, these vector elements are independent of $\epsilon$. This means, that in the limit $ \epsilon_{\backslash \backslash} = \epsilon_{\backslash} \rightarrow 0$ we approach a fixed steady state that is not the infinite temperature state $\Ket{I}$.   Moreover, the magnitude of this deviation from the thermal state is dictated primarily by the scale given by $1/r^{(2)}$ for which the term $1/r_1^{(2)}$  is the largest.

This outcome runs counter to typical perturbative statements where, as the small parameter tends to zero, we approach the bare un-perturbed state (in this case $\Ket{I}$). Recall however that, to avoid dealing with the massive degeneracy of the commutator $\mathbb{H}$, we also allowed the small parameter $\epsilon$ to enter into the bare Liouvillian. In this iterative construction then, we do not necessarily expect that each successive iteration will result in contributions that scale according to some positive power of $\epsilon$.  Indeed, one expects that further iterations would eventually lead to additional corrections in other $s$-even sectors that, similarly to the explicit first iterative correction above, do not vanish as $\epsilon \rightarrow 0$.

\section{The Liouvillian Gap}\label{sect:gap_results}
The next feature that we explore is the Liouvillian gap,  which one can consider as a key indicator of relaxation times towards the NESS~\cite{Dudzinski2000, Nagy2002, deGier2006, Kessler2012}. As shown in Eq. \ref{eq:L_gap}, this is defined $\mathcal{E}_{\text{gap}}  \equiv -\text{Re}(\mathcal{E}_{1})$, where $\mathcal{E}_{1}$ is the eigenvalue of $\mathcal{L}$ with the largest non-zero real component. For a review of gap behaviour in a variety of related models see~\cite{Znidaric2015}. Generically in such studies the key indicator is how the Liouvillian gap scales as a function of the system size, $N$, e.g. $\mathcal{E}_{\text{gap}}  \sim N^{-z}$ where the dynamical exponent $z$ depends on the particular model studied. 

\subsection{Emergence of an Open Gap from XY Anisotropy and Bulk Dissipation}
Utilising a convenient basis for the Liouvillian super-operator (Sec.~\ref{sect:OQ_basis}), we find that the gap for this system can be obtained via a MPS based approach~\cite{Orus2008, Prosen2009, Joshi2013}.  Moreover we find that, in this limit, the full Liovillian gap is closely shadowed by the gap obtained by restricting to the $s=2$ sector only - the  $\mathcal{E}_n^{(2)}$ gap used in the last section. Analysing the scaling of $s=2$ sector we find that it, and therefore the full Liouvillian gap scale as
\begin{equation}
	\mathcal{E}_{\text{gap}}  \sim f(\delta,h_{z}) + \mathcal{O}(N^{-1}),
\end{equation} 
where $f(\delta,h_{z})$ is non-zero when $|\delta|>0$. This implies that the relaxation time is finite in the thermodynamic limit when $|\delta| > 0$, since in this case the gap remains non-zero. This non-zero gap is not present in either XX + TASEP \cite{Temme2012}, XX + symmetric simple exclusion process (SSEP) \cite{Eisler2011}  or XY + boundary driving \cite{Prosen2009} models. As such we can infer that it is a consequence of combining both an XY anisotropy and bulk stochastic hopping. The precise functional form of the gap function $f(\delta, h_{z})$ for different types of dissipation,  including TASEP, remains an interesting question which we explore in future work~\cite{Kavanagh2021b}.

\subsection{MPS obtained $\mathcal{E}_{\text{gap}} $ versus $\mathcal{E}^{(2)}_{1}$  }\label{sect:TLGap}
Our key claims on the scaling of the gap are based on the assertion that, in the weak classical limit, the full Liouvillian gap can be estimated by only solving the $s=2$ sub-block.  Our key tool here is a  MPS calculation where we can effectively project out the steady state from the  variational algorithm. Here we exploit the structure that the Liouvillian super-operator takes in the so-called canonical Majorana representation (see Fig.~\ref{fig:Lblocks}), specifically using the fact that the $s=0$ block is only connected to the $s=2$ block via a single off-diagonal block, $\mathcal{L}^{(2,0)}$. This allows one to project out the steady state from the MPO that represents the full Liouvillian operator, while leaving all other eigenvalues unaffected.
  
\begin{figure}
\centering
\includegraphics[width=0.42\textwidth]{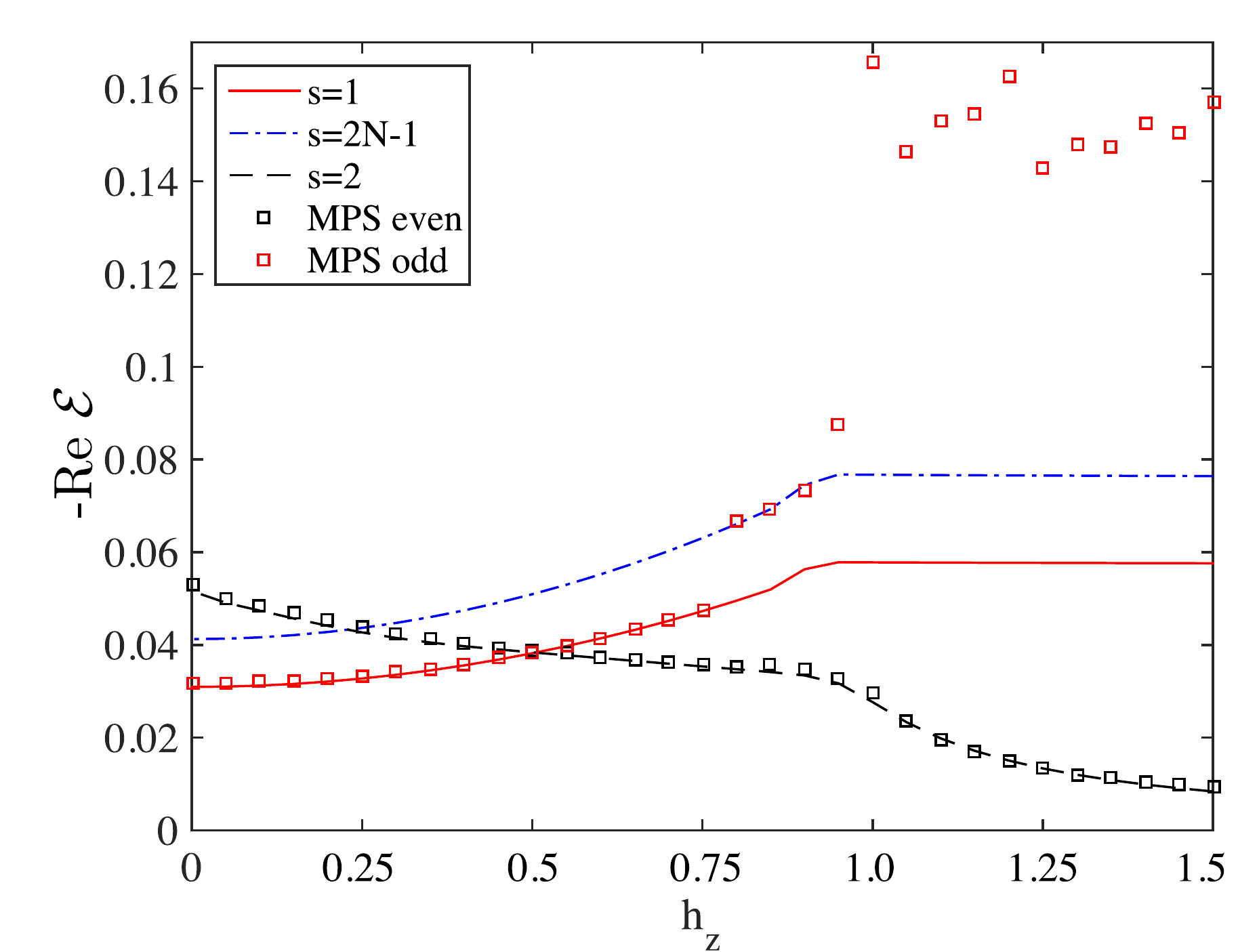}
\caption{Comparison of projection and MPS methods for a line cut at: $\alpha=0.1, \beta=0.3, \delta=0.7, \epsilon =0.1$  and $N=30$. A low virtual bond dimension ($\chi=20$ in this case) can be used to estimate gapped low lying states in both sectors by adding a weighted parity operator to $\mathcal{L}$. The even-sector gap can be estimated directly due to the specific form the Liouvillian takes in the canonical basis,  which means that one can decouple the $s=0$ block without affecting any other eigenvalues. }
\label{fig:MPScompare}
\end{figure}

In Fig.~\ref{fig:MPScompare} we compare the results from $ \mathcal{E}^{(2)}_{1}$ with the eigenvalues obtained from a full MPS treatment of a system of $N=30$ and see excellent agreement right across the phase diagram.  In App.~\ref{app:BPT} we also detail a perturbative argument for why these values are so close, using the  Rayleigh-Schr\"{o}dinger non-Hermitian formulation \cite{Sternheim1972} of the TXY-TASEP system. A synopsis of this calculation is that in the small $\epsilon$ regime, we can consider $s$-blocks as only being weakly connected to their $(s \pm 2)$-block neighbours.  Here spectrum $\mathcal{E}_{\text{gap}}$ can be expanded as
\begin{equation}
	\mathcal{E}_{\text{gap}} = \mathcal{E}^{(2)}_{1} + \mathcal{E}^{(2)\prime}_{1} + \mathcal{E}^{(2) \prime \prime}_{1}  + \dots,
\end{equation}
where $\mathcal{E}^{(s)}_{i}$ is the $i^{\text{th}}$ eigenvalue from the $s$ diagonal block and $\mathcal{E}^{(s) \prime}_i$ and $\mathcal{E}^{(s) \prime\prime}_i$ are the first and second order corrections. Crucially, one finds that the first order correction $\mathcal{E}^{(2)\prime}_{1}$ is zero and that the second order correction is much smaller than the zeroth order estimate, and typically scales as $\epsilon^{p}$ where $p > 2$, see App.~\ref{app:BPT}.

\subsection{Analysis of the $s=2$ Spectrum}
\begin{figure}
	\centering
	\includegraphics[width=.45\textwidth,height=0.36\textwidth]{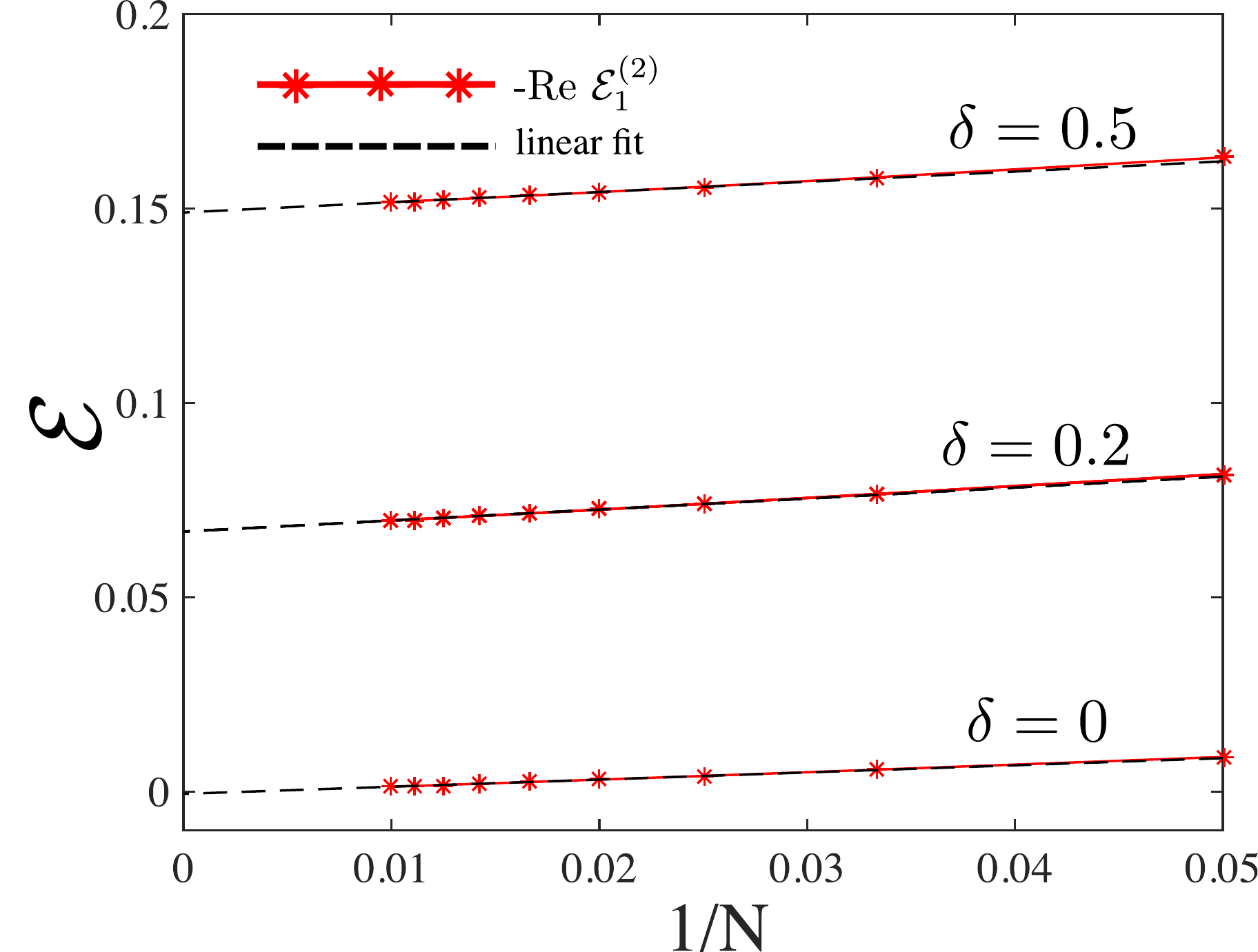}
	\includegraphics[width=.48\textwidth,height=0.36\textwidth]{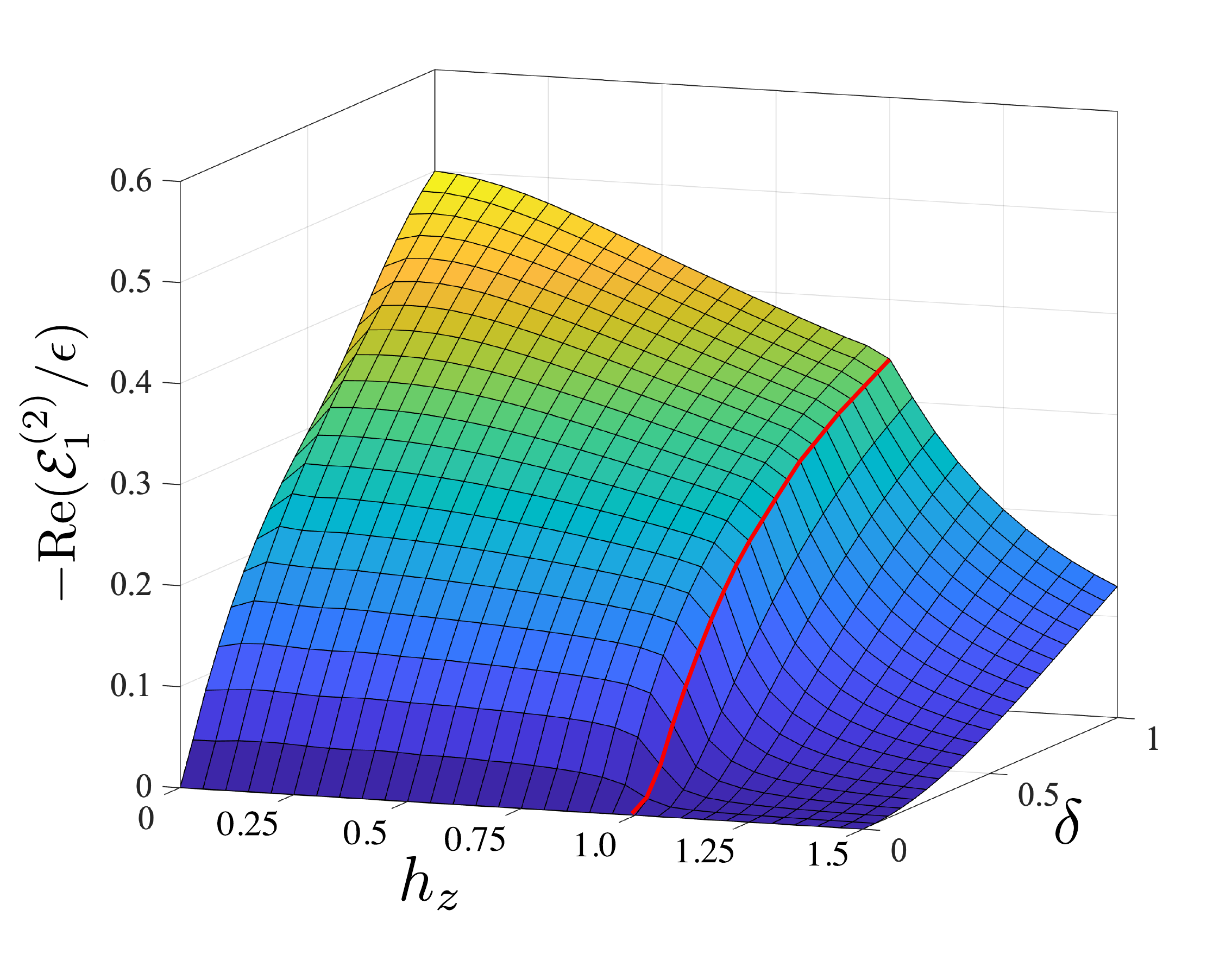}
	\caption[]{(Color online) [Top] Spectral gap scaling with $\alpha=0.1$, $\beta=0.3$, $\epsilon =0.1$, $h_z=0.5$. A nonzero $\delta$ introduces a persistent gap in the $N \rightarrow \infty$ limit. [Bottom] A scan of the projected $N \rightarrow \infty$ limit. The character of the gap changes when one traverses the quantum phase transition at $h_{z} = 1$ (red line). We note that the quantity plotted in the bottom figure is precisely $r^{(2)}_{1}$ of Sec.~\ref{sect:perturbation_NESS}.}
	\label{fig:gap_scaling}
\end{figure}

In the weak classical limit we can use $\mathcal{E}^{(2)}_{1} $ now as a proxy for the full gap and more fully analyse the parameter space of the model and assess its scaling as a function of system size, see Fig.~\ref{fig:gap_scaling}.  Our main result is that, in the the thermodynamic limit $N \rightarrow \infty$, the gap $\mathcal{E}_{\text{gap}} \rightarrow f(\delta,h_{z})$ remains open if the anisotropy parameter is non-zero.  However the dependence $\mathcal{E}_{\text{gap}} $ has on $\delta$ also relies strongly on the magnetic field parameter, with clear differences occurring between the different quantum phases of the Hamiltonian.  

When $\delta = 0$ we find that $f(0,h_{z}) = 0$ and thus $\mathcal{E}_{\text{gap}}  \sim N^{-1}\xrightarrow[ ]{N\rightarrow \infty} 0$.  This value is completely unaffected by changes in magnetic field $h_{z}$, as a result of a Lindblad symmetry present, see e.g. \cite{Albert2014}.  However, for non zero $\delta$  and when $|h_{z}| < 1$ (where the underlying Hamiltonian has a topological gap and boundary zero-energy modes) the Liouvillian gap develops linearly with $\delta$ (the superconducting order parameter in the fermionic picture). On the other hand where $|h_{z}| > 1$, and the system Hamiltonian is non-topological and the gap develops $\propto \delta^2$. This smaller gap means that perturbations to the thermal state are far more dramatic in this quantum regime. For a discussion on the odd sector blocks $s=1$ and $s=2N-1$ see App.~\ref{app:oddevengaps} and App.~\ref{app:oddspectrum}.

\subsection{Relaxation Rate Compared to Related Models}
	The interpretation of the gap as an inverse relaxation time leads one to consider the scaling of the gap with system size. If one has an inverse relation between the gap and the system size then in the large $N$ limit the system will not relax to the steady state in finite time. As such one often aims to determine the dynamical exponent, $z$, in the scaling relation $\mathcal{E}_{\text{gap}} \sim N^{-z}$. If $z = 0$, the longest relaxation times for the dynamics remain finite in the thermodynamic limit, while if $z > 0$ they diverge. 

	Generically, the dynamical exponent depends on a variety of factors from the model in question. The gap scaling of our model has been found in certain special cases. It has been determined analytically \cite{Prosen2008b, Prosen2008} for the quantum XY model, with bi-directional dissipation on boundary sites only, that the gap scales as $\mathcal{E}_{\text{gap}}  \sim N^{-3}$ everywhere except for at $|h_{z}| = 1- \delta^2 $ where the gap closes more rapidly as: $\mathcal{E}_{\text{gap}}  \sim N^{-5}$. This can be contrasted with the TASEP gap scalings which differ depending on the phase of the classical model. There, one finds by various approaches \cite{deGier2005, deGier2006, Znidaric2015, Dudzinski2000, Nagy2002} gap scalings of $\mathcal{E}_{\text{gap}}  \sim N^{0}$ in the LD/HD phases, $\mathcal{E}_{\text{gap}}  \sim N^{-3/2}$ in the MC phase, and $\mathcal{E}_{\text{gap}}  \sim N^{-2}$ on the critical line, where $\alpha = \beta$.

A relatively generic bound for the gap scaling of $\mathcal{E}_{\text{gap}}  \sim N^{-1}$ can be found for systems with only boundary dissipation~\cite{Znidaric2015} which is a component of the model studied in this paper. However, the existence of a finite gap for appropriate Hamiltonian parameters places the TXY-TASEP outside of the scope of these results and indeed also outside the scope of integrable systems results~\cite{Ziolkowska2020, deLeeuw2021}. One can draw the conclusion that both bulk and boundary dissipation together are necessary for a non-vanishing gap in all phases of the TASEP.

\section{Summary and Conclusion}\label{sect:conclusion}
In this work we have explored the transverse XY TASEP system, showing how the interplay between XY anisotropy and transverse field affect both the non-equilibrium steady state and the gap that separates it from the rest of the Liouvillian spectrum.

An interesting aspect to the model is the ability to tune between different steady states that derive key properties from the underlying quantum phase. These quantum effects are most profound in the parameter spaces of low magnetic field ($h_{z} < 1$) where the XY terms opens a Liouvillian gap that is approximately linear in the anisotropy $\delta$. On the other hand, in the regimes associated with high transverse field ($h_{z} > 1$) we see that the steady state essentially reverts to the something like the purely stochastic NESS, mimicking the scenario also found with no XY anisotropy, albeit with a gap now proportional to $\delta^2$.

The low field deviations from the classical NESS, most pronounced in the TASEP low and high density regimes, can be understood by viewing the XY anisotropy $\delta$ as a source of pair creation/annihilation which seeks to drive the system towards half filling, and pin the energy expectation value to energies close to the centre of the many-body spectrum. The high magnetic effect reduces this anisotropic drive toward half filling allowing the particle densities to be largely determined by the classical boundary driving. This coincides with NESS energy expectation values drifting towards the extremes of the Hamiltonian many-body spectra. 

Our observations of a constant gap show that the TXY-TASEP system constitutes what is called a rapidly mixing system. In this respect the canonical Majorana basis provides an intuitive way to understand this in terms of the perturbations to the maximally mixed state and the gaps found in successive even-parity excitation number blocks. In cases where the even quasi-particle gap is constant we can argue that, in the weak classical limit, that successive perturbations will decay order-by-order. On the other hand, when the gap closes with system size, or is at least very small (as in the large transverse field limit), one expects that successive perturbations to the maximally mixed-state will not completely decay, so that the resulting NESS remains very different. 

The mechanism we describe shows how the quantum system can be used to rapidly switch between radically different steady states, either by tuning $\delta$ or the the transverse field $h_{z}$. A natural question to ask then is what types of pre-determined states that can be easily prepared in this fashion? Moreover, can TXY-TASEP be a template from which one can develop such schema?  In this paper we argue that this so-called rapidly mixing aspect is due to both the bulk driving and XY anisotropy (due to the lack of similar effects for boundary driven only or XX systems). An interesting question is if the XY model gives similar results for other types of bulk driving/dissipation? Other work in this area suggest that this may indeed be a general feature. In Joshi et. al. \cite{Joshi2013} the XY model with bulk dissipation showed distinctly different behaviours of the steady-state negativity in the different quantum regimes. Moreover, they also observed rapidly decaying correlations in the topological/ferromagnetic regime. This is consistent with a robust Liouvillian gap and thus indicates that XY systems, together with bulk Lindblad operators generally, may prove a promising avenue for rapid state preparation. 

\begin{acknowledgments}
We acknowledge Ian Hughes for inspiring discussions in the early stages of this work. K.K., S.D., and G.K. acknowledge Science Foundation Ireland for financial support through Career Development Award 15/CDA/3240. G.K. was also supported by a Schr{\"o}dinger Fellowship.  J.K.S. was supported through SFI Principal Investigator Award 16/IA/4524. 
\end{acknowledgments}

\bibliographystyle{apsrev4-1}
\bibliography{refs}

\newpage
\appendix

\section{TASEP Embedded in a Quantum Spin Chain}\label{app:discrete_TASEP}
The classical TASEP dynamics are usually described as a stochastic discrete-time update rule \cite{Derrida1992}. To update the configuration $\vec{n}_t$ to the configuration $\vec{n}_{t + \Delta t}$ a time-step later, we randomly choose an integer from the set $\{ 0,1,...,N \}$ with a uniform distribution, i.e., each integer has a probability $1/(N+1)$ of being selected.

\begin{enumerate}
\item \label{step:hop} If the outcome is $i \in \{1,...,N-1\}$, and if $n_i = 1$ and $n_{i+1} = 0$, we hop the particle from site $i$ to site $i+1$ with the probability $\gamma \Delta t$.
\item \label{step:on} If the outcome is $i = 0$, and if $n_1 = 0$, then we should introduce a particle at site $i=1$ with the probability $\alpha \Delta t$.
\item \label{step:off} If the outcome is $i = N$, and if $n_N = 1$, then we should remove the particle at site $i = N$ with the probability $\beta \Delta t$.
\end{enumerate}

We would like to represent this discrete-time state update rule as an operation on the $N$-qubit classical state, that preserves the classical nature of the state, i.e., takes a diagonal density matrix to another diagonal density matrix. To do this, we first consider the quantum operation that represents step \ref{step:hop} above, i.e., the hopping of a particle from site $i$ to site $i+1$. This can be implemented with the operation $\hat{\rho} \to \Lambda_i^\text{hop}[\hat{\rho}]$, where: 
\begin{equation} 
	\Lambda_i^\text{hop}[\hat{\rho}] = \sum_{j=0}^1 \hat{K}_i^{(j)} \hat{\rho} \hat{K}_i^{(j) \dagger} , \quad i \in \{ 1,2,\hdots,N-1 \}, 
\end{equation} 
for the Kraus operators: 
\begin{eqnarray} 
	&&\hat{K}_i^{(0)} \equiv \ket{0_i 0_{i+1}} \bra{0_i 0_{i+1}} + \ket{0_i 1_{i+1}} \bra{0_i 1_{i+1}} + \nonumber \\ 
	&&\sqrt{1-\gamma\Delta t} \ket{1_i 0_{i+1}} \bra{1_i 0_{i+1}} + \ket{1_i 1_{i+1}} \bra{1_i 1_{i+1}}, \label{eq:K_i_0} \\ 
	&&\hat{K}_i^{(1)} \equiv \sqrt{\gamma\Delta t} \ket{0_i 1_{i+1}} \bra{1_i 0_{i+1}} . \label{eq:K_i_1} 
\end{eqnarray} 
Intuitively, the Kraus operator $\hat{K}_i^{(1)}$ hops a particle from site $i$ to site $i+1$, with probability $\gamma \Delta t$, only if site $i$ is occupied and site $i+1$ is unoccupied, i.e., it implements step \ref{step:hop}. The Kraus operator $\hat{K}_i^{(0)}$ leaves the system unaffected in all other cases. It is easily checked that $\hat{K}_i^{(0) \dagger} \hat{K}_i^{(0)} + \hat{K}_i^{(1) \dagger} \hat{K}_i^{(1)} = \hat{\mathds{1}}$, making this is a well-defined, probability preserving quantum operation. It is also easily checked that this operation preserves the classical nature of a state, since it takes any diagonal density matrix $\hat{\rho}$ to another diagonal density matrix $\hat{\rho}$. 

Similarly, the quantum operation that represents steps \ref{step:on} and \ref{step:off} above are $\hat{\rho} \to \Lambda^\text{on}[\hat{\rho}]$ and $\hat{\rho} \to \Lambda^\text{off}[\hat{\rho}]$, respectively, where: 
\begin{equation} 
	\Lambda^\text{on}[\hat{\rho}] = \sum_{j = 0}^1 \hat{K}_\text{on}^{(j)} \hat{\rho} \hat{K}_\text{on}^{(j) \dagger} , \quad \Lambda^\text{off}[\hat{\rho}] = \sum_{j = 0}^1 \hat{K}_\text{off}^{(j)} \hat{\rho} \hat{K}_\text{off}^{(j) \dagger}, 
\end{equation} 
for the Kraus operators: 
\begin{eqnarray} 
	\hat{K}_\text{on}^{(0)} &\equiv& \sqrt{1 - \alpha\Delta t} \ket{0_1}\bra{0_1} + \ket{1_1}\bra{1_1}, \\ \quad \hat{K}_\text{on}^{(1)} & \equiv& \sqrt{\alpha\Delta t} \ket{1_1}\bra{0_1}, \\ 
	\hat{K}_\text{off}^{(0)} &\equiv& \ket{0_N}\bra{0_N} + \sqrt{1 - \beta\Delta t} \ket{1_N}\bra{1_N}  ,\\ \quad \hat{K}_\text{off}^{(1)} &\equiv& \sqrt{\beta\Delta t} \ket{0_N}\bra{1_N} . 
	\end{eqnarray} 
	Again, it is straightforward to check that $\hat{K}_\text{on}^{(0) \dagger} \hat{K}_\text{on}^{(0)} + \hat{K}_\text{on}^{(1) \dagger} \hat{K}_\text{on}^{(1)} = \hat{\mathds{1}}$ and $\hat{K}_\text{off}^{(0) \dagger} \hat{K}_\text{off}^{(0)} + \hat{K}_\text{off}^{(1) \dagger} \hat{K}_\text{off}^{(1)} = \hat{\mathds{1}}$, and also that these operations preserve the classical (i.e. diagonal) nature of a state $\hat{\rho}$.

Implementing each of these possibilities with the uniform probability $1/(N+1)$ gives the full quantum operation representing the discrete-time state update: 
\begin{eqnarray} \hat{\rho}(t + \Delta t) &=& \Lambda [\hat{\rho}(t)] = \frac{1}{N+1}\Lambda^\text{on}[\hat{\rho}(t)] \\ &+& \frac{1}{N+1}\Lambda^\text{off}[\hat{\rho}(t)] + \frac{1}{N+1} \sum_{i=1}^{N-1}\Lambda_i^\text{hop}[\hat{\rho}(t)] . \nonumber
\end{eqnarray}  

We can find the classical continuous-time master equation in the $\Delta t \to 0$ limit of the discrete dynamics above. First, we focus on the hopping operation $\Lambda_i^\text{hop}$. For this operation alone, the master equation is found as: 
\begin{equation} 
	\lim_{\Delta t \to 0}\frac{\Lambda_i^\text{hop}[\hat{\rho}(t)] - \hat{\rho}(t)}{\Delta t} = \gamma \mathcal{L}[\hat{\sigma}_i^{-}\hat{\sigma}_{i+1}^{+}](\hat{\rho}(t)) , \end{equation} 
	where: $\mathcal{L}(\hat{O})[\hat{\rho}] \equiv \hat{O}\hat{\rho}\hat{O}^\dagger - \frac{1}{2}\hat{O}^\dagger \hat{O}\hat{\rho} - \frac{1}{2}\hat{\rho}\hat{O}^\dagger \hat{O}$. Similarly, for the $\Lambda^\text{on}$ and $\Lambda^\text{off}$ processes, we have: 
\begin{eqnarray} 
		\lim_{\Delta t \to 0}\frac{\Lambda^\text{on}[\hat{\rho}(t)] - \hat{\rho}(t)}{\Delta t} &=& \alpha \mathcal{L}[\hat{\sigma}_{1}^{+}](\hat{\rho}(t)),  \\ 
		\lim_{\Delta t \to 0}\frac{\Lambda_i^\text{off}[\hat{\rho}(t)] - \hat{\rho}(t)}{\Delta t} &=&  \beta \mathcal{L}[\hat{\sigma}_{N}^{-}](\hat{\rho}(t))  , 
\end{eqnarray} 
respectively. Combining each of these gives the TASEP continuous-time master equation: 
\begin{eqnarray} 
	\frac{d}{dt}\hat{\rho}(t) &=& \frac{1}{N+1} \left( \alpha \mathcal{L}(\hat{\sigma}_{1}^{+}) + \beta \mathcal{L}(\hat{\sigma}_{N}^{-})\right) \\ &+&  \frac{1}{N+1}  \left( \gamma \sum_{i=1}^{N-1} \mathcal{L}(\hat{\sigma}_i^{-}\otimes\hat{\sigma}_{i+1}^{+}) \right) [\hat{\rho}(t)]  \nonumber. 
\end{eqnarray} Finally, rescaling the rates $\alpha$, $\beta$, $\gamma$ by a factor of $N+1$ gives Eq.~\ref{eq:cl_TASEP} in the main text, where $\gamma$ is additionally set to 1.

\section{Block Perturbation Theory}\label{app:BPT}
The structure of the Lindblad operator in the canonical basis allows one to see why, in the weak classical limit, one can typically focus on the extremum blocks $s=0,1,2$ and $2N-1$ to understand the gap scaling.  Starting in the canonical basis, we generalise our previous notation and also label the block matrices according to the  excitation number blocks that they connect. For example $\mathcal{L}^{(0,2)}$ is the block-matrix that connects the $0^{\text{th}}$ and $2^{\text{nd}}$ excitation number blocks, whereas, like before, $\mathcal{L}^{(n)}$ labels the $n$-excitation number diagonal.

We wish to understand how the coupling to other blocks affects the energies of a particular block and so employ a Rayleigh-Schr\"{o}dinger non-Hermitian formulation problem \cite{Sternheim1972}, which proceeds very similar to the Hermitian counterpart. We consider the diagonal blocks as our unperturbed system
\begin{equation}
	\mathcal{L}_0 = \sum_{ s  \in even}  [\mathcal{L}^{(s)}  ] =  \sum^{2N}_{s  \in \text{ even}}  \epsilon_{\backslash}   \mathbb{L}^{(s)}_{\backslash} -i \lambda \mathbb{H}^{(s)}
\end{equation}  
and the perturbation as the off diagonal complement
\begin{align}
	\mathcal{L}_1 &=& \mathcal{L}-\mathcal{L}_0 =  \sum_{s  \in \text{ even}}   \mathcal{L}^{(s,s+2)} + \mathcal{L}^{(s+2,s)} \\
	&=&  \epsilon_{\backslash \backslash } \sum_{s  \in \text{ even}}   \mathbb{L}^{(s,s+2)} + \mathbb{L}^{(s+2,s)}
\end{align}

The left and right eigenvectors $\Ket{\tilde{v}^{(n)}_{i}}$ and $\Ket{v^{(n)}_{i}}$ are those which diagonalise the diagonal blocks $\mathcal{L}^{(n)}$
\begin{equation}
	\Bra{\tilde{v}^{(l)}_{i}} \mathcal{L}^{(n)} \Ket{v^{(m)}_{j}}=  \mathcal{E}^{(n)}_{i} \delta_{ij}\delta_{nm}\delta_{ln},
\end{equation}

Starting with one of the zeroth order states which we obtained by diagonalising one of the diagonal blocks $\mathcal{L}^{(n)}$ we wish to understand how the addition of the off diagonal blocks perturb this energy :
\begin{equation}
	\mathcal{E}^{(n)}_{i,\text{exact}} =   \mathcal{E}^{(n)}_{i} + \mathcal{E}^{(n)\prime}_{i} +  \mathcal{E}^{(n)\prime\prime}_{i} +... 
\end{equation}
The first order correction $\mathcal{E}^{(n)\prime}_{i} $ can be easily seen to vanish simply because $\mathcal{L}_1$ does not connect any block to itself
\begin{equation}
	\mathcal{E}^{(n)\prime}_{i} = \Bra{\tilde{v}^{(n)}_{i}} \mathcal{L}_1 \Ket{v^{(m)}_{j}}=0.
\end{equation}
The leading correction to the eigenvalue can thus only occur at second order or higher.  Generally, the second order correction can be written as
\begin{equation}
	\mathcal{E}^{(n)\prime\prime}_i= \sum_{j,m} \frac{ \Bra{\tilde{v}^{(l)}_{i}}  \mathcal{L}_1\Ket{v^{(n)}_{i}} \Bra{\tilde{v}^{(l)}_{i}}  \mathcal{L}_1 \Ket{v^{(n)}_{i}}} { \mathcal{E}^{(n)}_{i}-\mathcal{E}^{(n)}_{j}}.
\end{equation}

For the steady state $\ket{v^{(0)}}$ with $\mathcal{E}^{(0)} =0$ we see that, because $\mathcal{L}^{(0,2)}= 0$, there can be  no higher order corrections to this eigenvalue (as one would expect).  In our MPS calculations the same feature can be used to decouple the steady state from the even parity sector and allows us to converge variationally on the first even-parity excited state  above the gap. 

Our primary focus here is to understand the energy scaling of states from the $s=2$ excitation-number block on a perturbative level. As $\mathcal{L}^{(0,2)} = 0$ we only have to consider perturbative paths that connect to the $s=4$ block and thus:
\begin{equation}
	\mathcal{E}^{(2)\prime\prime}_{i} = \sum_{j} \frac{ \Bra{\tilde{v}^{(2)}_{j}}\mathcal{L}^{(2,4)} \Ket{v^{(4)}_{j}} \Bra{\tilde{v}^{(4)}_{j}} \mathcal{L}^{(4 ,2)} \Ket{v^{(2)}_i} } { \mathcal{E}^{(2)}_{i}-\mathcal{E}^{(4)}_{j}}.
\end{equation}
The analysis above is fairly conventional. However there is one anomaly in that we have hidden the small parameter $\epsilon$ in both the diagonal and off diagonal blocks. Thus, we expect the parameter $\epsilon$ to appear in both numerator and denominator of the second order expansion above. For the off-diagonal terms the contribution there is an overall $\epsilon^2$ factor in each of the $\mathcal{L}$ operators. However, we may also expect some $\epsilon$ contributions in both the zeroth order eigenstates and in the real part of the eigenvalues appearing in the denominator. 

\begin{figure}
\centering
\includegraphics[width=.48\textwidth]{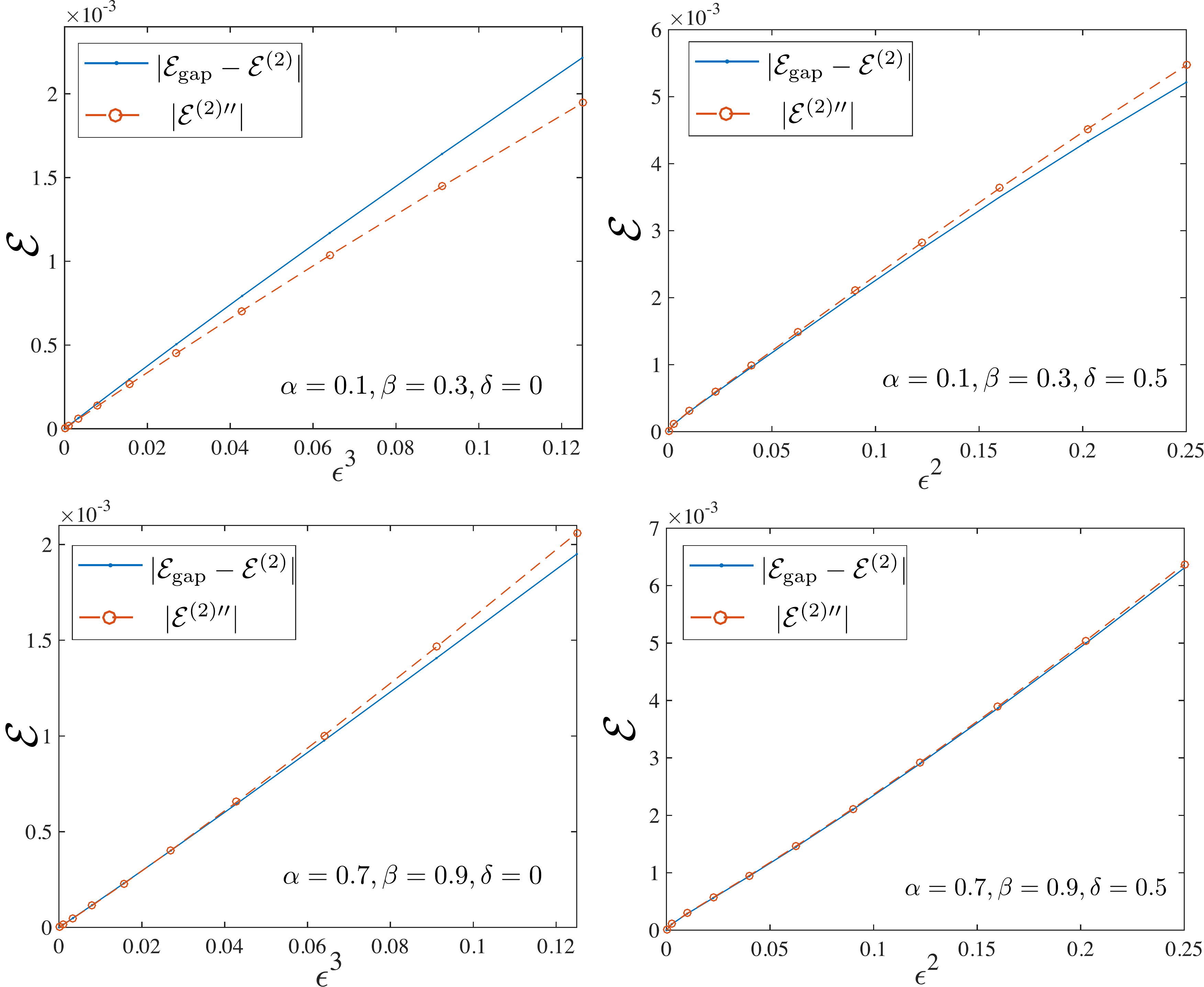}
\caption{The difference between the $\mathcal{E}_{\text{gap}}$ and $\mathcal{E}^{(2)}_1$ along with the  second order correction $\mathcal{E}^{(2) \prime\prime}$ for a system size of $N=7$ with $\delta=0$ and $0.5$ with boundary driving $(\alpha,\beta)$ of $(0.1,0.3)$ and $(0.7,0.9)$. }
\label{fig:eps_scaling}
\end{figure}

If the imaginary part of the denominator is small with respect the real part then we see an $\epsilon^{-1}$ contribution occurring from these terms. In practice, however, we see that most of the weight of the occurs in the opposite limit where the $\epsilon^{-1}$ contribution is negligible. Indeed, we have found that this $\epsilon^{-1}$ scaling is compensated for via the $\epsilon$ dependence within the eigenstates themselves, leaving a net scaling of $\epsilon^r$ with $r\ge2$. In Fig.~\ref{fig:eps_scaling} we compare the $\mathcal{E}_{\text{gap}}$ with the zeroth  $\mathcal{E}_{2}^{[0]}$  estimate along with second order $\mathcal{E}_{2}^{[2]}$ correction. We see that for $\delta = 0$ the correction seems to actually scale close to $\epsilon^3$ while for $\delta=0.5$ the scaling is closer to  $\epsilon^2$ . 

\begin{figure}
\centering
\includegraphics[width=.48\textwidth]{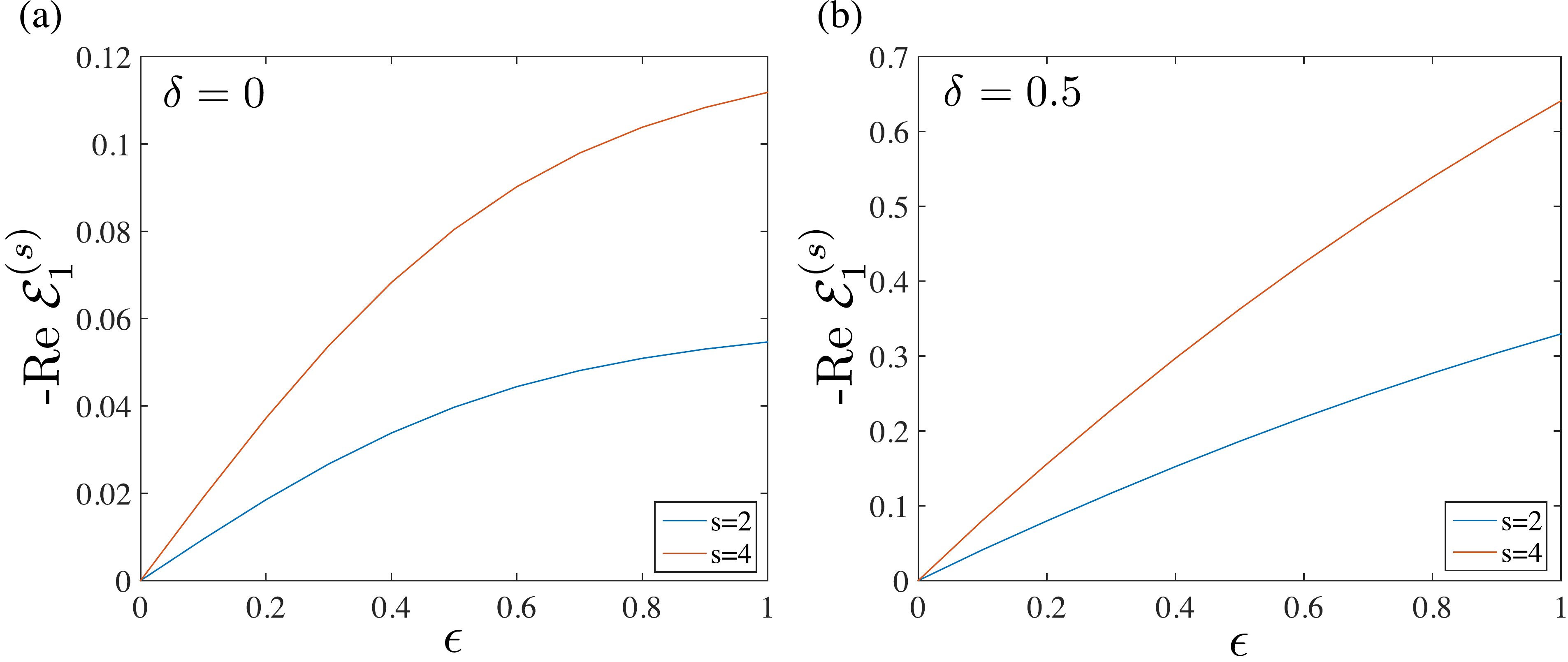}
\caption{The minimum values of$|\text{Re} \mathcal{E}^{(2)}|$ and $|\text{Re} \mathcal{E}^{(4)}|$ for $N = 16$, $\alpha = 0.7$, $\beta = 0.9$, and $h_{z} = 1$. In the small $\epsilon$ limit these eigenvalues are both proportional to $\epsilon$ and we have $|\mathcal{E}^{(0)}_1| \approx 2 |\mathcal{E}^{(2)}_1|$.  }
\label{fig:S2S4_scaling}
\end{figure}

One last question remains; how can we be sure that the real gap magnitude of $\mathcal{E}^{(2)}_{1}$ is smaller than that of $\mathcal{E}^{(2)}_{1}$? Here we proceed by observation.  In Fig.~\ref{fig:S2S4_scaling} we show the behaviour of the absolute value of eigenvalues that are closest to the steady state for a system size of $N=16$.  In this figure and in all other similar calculations we see that the the minimum even gap magnitude comes from the $s=2$ sector, and that it is approximately half that of the $s=4$ sector.  

\section{The Meaning of Even and Odd Sector Gaps}\label{app:oddevengaps}
	In the main text we distinguished between the maximum non-zero real eigenvalues from both even and odd parity sectors of the Liouvillian. It is worth discussing briefly what these eigenvalues represent. Firstly, we note that the block-diagonal structure (see Fig.~\ref{fig:Lblocks}) can be interpreted as excitation number conservation, which in the representation used here resembles magnetisation on a $2N$-site spin chain. Interactions and/or dissipation can break this symmetry but still allow excitation parity conservation. Parity then allows us to divide up the full $2^{2 N}$ dimensional space into two $2^{2N-1}$ dimensional spaces.  

	The even sector consists of operators that preserve the parity of a state. This includes density operators $\rho = \sum p_{\psi} \ket{\psi}\bra{\psi}$ where $\ket{\psi}$ have well defined parity. For this reason the even-sector gap is what determines the slowest relaxation rate towards the steady state. We have argued that this gap can be largely understood by focusing on the $s=2$ block of the basis rotated Liouvillian super-operator.

	The operator Hilbert space allows for more possibility than density matrices. The odd-sector of the super-operator $\mathcal{L}$, for example, consists of basis states that represent fermionic creation and annihilation operators \cite{Kells2015} and odd numbered products of them. By definition such operators would switch the parity of a state. For excitation number preserving systems suitable combinations of these single-particle operators ($\Gamma^{(1)}$-sector) are the quasi-particle excitations and by combining products of such operators one can generate more complicated $n$-particle excitation operators in the other excitation number blocks \cite{Kells2015}. Although this meaning is diluted if there is no longer excitation number symmetry, it is important to know where such states occur in order to distinguish them from the even sector gap. We will see again that the extremal sectors ($s=1$ and $s=2N-1$ in this case) allow us to predict the largest odd sector eigenvalue. 

\section{Spectrum of the Odd Sectors}\label{app:oddspectrum} 
In the canonical basis for the $s=1$  and $s=2N-1$ sub-blocks the elements from the commutator can be read directly from the adjacency-matrix used to define the quadratic Hamiltonian (see e.g. \cite{Goldstein2012, Kells2015}).  In these sub-blocks the terms from stochastic process occur only on the diagonal:
\begin{eqnarray}
	\non
	\mathcal{L}_{n,n} ^{ (1)} &=& -\epsilon [ 1/2+ \alpha -  (\alpha/2+ 1/4) (\delta_{n,2N-1}+\delta_{n,2N})  \\ 
	&& -(\beta/2+1/4) (\delta_{n,1} +  \delta_{n,2}) ], 
	\label{eq:A1}
\end{eqnarray}
\begin{eqnarray}
\non \mathcal{L}_{n,n} ^{(2N-1)} &=& -\epsilon [ 1/2+ \beta-  (\beta/2+ 1/4) (\delta_{n,2N-1}+\delta_{n,2N})  \\ 
&& -(\alpha/2+1/4) (\delta_{n,1} +  \delta_{n,2}) ]. 
\label{eq:A2nm1}
\end{eqnarray}

Setting, as throughout, the bulk stochastic hopping amplitude to $1$ and neglecting the boundary terms we see that for the $s=1$ ($s=2N-1$) sector the hop-on (hop-off) coefficient $\alpha$ ($\beta$) acts constantly throughout the bulk of the system and thus the largest real eigenvalues in each sector are effectively linearly dependent on these hop-on and hop-off rates. 

On top of this linear dependence, the imaginary components stemming from the Hamiltonian part of the Liouvillian also play a crucial role. In the topologically trivial phase ($|h_z| > J$) the bulk imaginary spectrum in the continuum limit behaves approximately as
\begin{equation}
	\text{Im} (E) =  \pm \sqrt{ (-h_{z} + J \cos (k))^2 + (\delta \sin(k)^2)}.
\end{equation}
In the  ferromagnetic/topological phase ($h_z < J$) the open system develops evanescent edge modes on the $\text{Im}(E)=0$ line. These modes are the so-called Majorana zero modes that have been studied extensively in recent years \cite{Kitaev2001, Fu2008, Lutchyn2010, Oreg2010, Stanescu2013}. In the limit that these zero-modes have a very long coherence length $\xi \propto J/\delta \gg 1$ (i.e. small $\delta$) we see that the associated real component saturates to the bulk value of $\sim -\epsilon (\gamma/2+ \alpha)$ or $-\epsilon ( \gamma/2+ \beta)$ see (\ref{eq:A1}) and (\ref{eq:A2nm1}). In the ferromagnetic limit ($\delta=1$ and $h_z=0$) the zero-modes are $\delta$-functions at sites $n=1$ and $n=N$ and thus the real components can be estimated as  $-\epsilon (1/2+ \alpha)/2,-\epsilon (1/2+ \beta)/2,-\epsilon (1/4+ \alpha -\beta/2),-\epsilon (1/4 +\beta- \alpha/2)$. When $\alpha$ and $\beta$ are both small these topological driven states slice through the even-sector gap to become closest to the steady state, see Fig.~\ref{fig:Lblocks} (c) and Figure \ref{fig:MPScompare}.

\end{document}